\documentclass[lettersize,journal]{IEEEtran}
\usepackage{lineno}
\usepackage{cite}
\usepackage{hyperref}
\usepackage{amsmath,amssymb,amsfonts}
\usepackage{amsmath}
\usepackage{amsthm}

\usepackage{algpseudocode}
\usepackage{graphicx}
\usepackage{textcomp}
\usepackage{xcolor}
\usepackage{graphicx}
\usepackage{float}
\usepackage{stfloats}
\usepackage{subfigure}
\usepackage{amsmath}
\usepackage{amsfonts,amssymb}
\usepackage{mathrsfs}
\usepackage{mathtools}
\usepackage{algorithm}
\usepackage{algorithmicx}
\usepackage{algpseudocode}
\usepackage{bm}
\usepackage{multirow}
\usepackage{array}
\usepackage{amssymb}
\usepackage{amsmath}
\usepackage{cite}
\usepackage{url}
\usepackage{xcolor}
\usepackage{cite,graphicx,amsmath,amssymb}
\usepackage{subfigure}
\usepackage{fancyhdr}
\usepackage{mdwmath}
\usepackage{mdwtab}
\usepackage{caption}
\usepackage{amsthm}
\usepackage{setspace}
\usepackage{bm}
\usepackage{algorithm}
\usepackage{algpseudocode}
\usepackage{mathtools}
\usepackage{dsfont}
\usepackage{bbm}
\usepackage{cases}
\usepackage{booktabs}

\theoremstyle{definition}
\newtheorem{theorem}{Theorem}

\newtheorem{lemma}{Lemma}

\newtheorem{corollary}{Corollary}

\makeatletter
\newcommand{\biggg}{\bBigg@{3}}
\newcommand{\Biggg}{\bBigg@{3.5}}
\makeatother
\begin{document}

\title{Secure Beamforming for Continuous Aperture Array (CAPA) Systems}
\author{
\author{
Mingjun Sun,~\IEEEmembership{Graduate Student Member,~IEEE,} Chongjun Ouyang,~\IEEEmembership{Member,~IEEE,}\\ Zhaolin Wang,~\IEEEmembership{Member,~IEEE,} Shaochuan Wu,~\IEEEmembership{Senior Member,~IEEE,} and Yuanwei Liu,~\IEEEmembership{Fellow,~IEEE}
\thanks{Mingjun Sun and Shaochuan Wu are with the School of Electronics and Information Engineering, Harbin Institute of Technology, Harbin 150001, China (e-mail: sunmj@stu.hit.edu.cn; scwu@hit.edu.cn).}
\thanks{Chongjun Ouyang is with the School of Electrical and Electronic Engineering, UCD College of Engineering and Architecture, University College Dublin, D04 V1W8, Ireland, and also with the School of Electronic Engineering and Computer Science, Queen Mary University of London, London, E1 4NS, U.K. (e-mail: chongjun.ouyang@ucd.ie).}
\thanks{Zhaolin Wang is with the School of Electronic Engineering and Computer Science, Queen Mary University of London, London E1 4NS, U.K. (e-mail: zhaolin.wang@qmul.ac.uk).}
\thanks{Yuanwei Liu is with the Department of Electrical and Electronic Engineering, The University of Hong Kong, Hong Kong (e-mail: yuanwei@hku.hk).}
}}

%

\maketitle

\begin{abstract}
Continuous aperture array (CAPA) is considered a promising technology for 6G networks, offering the potential to fully exploit spatial DoFs and achieve the theoretical limits of channel capacity. This paper investigates the performance gain of a CAPA-based downlink secure transmission system, where multiple legitimate user terminals (LUTs) coexist with multiple eavesdroppers (Eves). The system's secrecy performance is evaluated using a weighted secrecy sum-rate (WSSR) under a power constraint. We then propose two solutions for the secure current pattern design.
The first solution is a block coordinate descent (BCD) optimization method based on fractional programming (FP), which introduces a continuous-function inversion theory corresponding to matrix inversion in the discrete domain. This approach derives a closed-form expression for the optimal source current pattern. Based on this, it can be found that the optimal current pattern is essentially a linear combination of the channel spatial responses, thus eliminating the need for complex integration operations during the algorithm's optimization process. The second solution is a heuristic algorithm based on Zero-Forcing (ZF), which constructs a zero-leakage current pattern using the channel correlation matrix. It further employs a water-filling approach to design an optimal power allocation scheme that maximizes the WSSR. In high SNR regions, this solution gradually approaches the first solution, ensuring zero leakage while offering lower computational complexity. Simulation results demonstrate that: 1) CAPA-based systems achieve better WSSR compared to discrete multiple-input multiple-output (MIMO) systems. 2) The proposed methods, whether optimization-based or heuristic, provide significant performance improvements over existing state-of-the-art Fourier-based discretization methods, while considerably reducing computational complexity.
\end{abstract}

\begin{IEEEkeywords}
Continuous aperture array (CAPA), fractional programming, optimal source current patterns, weighted secrecy sum-rate.
\end{IEEEkeywords}

\section{Introduction}
\IEEEPARstart{I}{n}
the ever-evolving landscape of wireless communication, achieving higher capacity and enhanced reliability remains a central focus. MIMO technology has played a transformative role in meeting these demands \cite{ref1}, evolving from its foundational application in 3G networks to advanced configurations like Massive MIMO \cite{ref2} in 5G and Gigantic MIMO \cite{ref3} envisioned for 6G. However, despite its successes, MIMO systems, which rely on spatially discrete antenna arrays, face intrinsic limitations due to physical constraints on antenna density and aperture size. These limitations hinder their ability to fully exploit the spatial DoFs for next-generation communication systems.

In recent years, various new high-density antenna paradigms, such as holographic MIMO \cite{ref4}, \cite{ref5} and large intelligent surfaces (LIS) \cite{ref6}, \cite{ref7}, have emerged to address these challenges. In this paper, we collectively refer to the ultimate version of these technologies as \textit{continuous aperture arrays} (CAPAs), which have the potential to achieve ultimate array gains \cite{ref8}, \cite{ref9}.
CAPAs offer a revolutionary approach by replacing discrete antenna arrays with a continuous electromagnetic (EM) radiating surface. Unlike conventional discrete MIMO systems, CAPAs enable the continuous distribution of source currents across the aperture, leveraging the full spatial DoFs \cite{ref29} while simplifying hardware requirements. Specifically, CAPAs require fewer radio frequency (RF) chains than discrete antenna systems for the same number of multiplexed signals, resulting in improved energy efficiency and scalability \cite{ref10}. Additionally, their ability to precisely control the amplitude and phase of current distributions positions CAPAs as a powerful solution for achieving enhanced communication performance.

Early studies on CAPAs focused on fundamental performance metrics, including the analysis of DoFs \cite{ref6, ref11, ref12}, and channel capacity \cite{ref13, ref14, ref15,ref16} under both line-of-sight (LoS) and multipath scattering environments. The work in \cite{ref6} demonstrated that LIS achieved spatial DoFs greater than one, even under strong LoS conditions, significantly enhancing communication capacity. The study in \cite{ref11} introduced a rigorous method for calculating the communication DoFs between two arbitrarily shaped and positioned volumes by solving eigenvalue problems, without relying on planar surface assumptions or paraxial approximations. Furthermore, in \cite{ref12}, the authors developed a spatial bandwidth-based analytical framework that provided closed-form approximations for achievable spatial DoFs in LoS channels with large-scale antenna arrays, considering the effects of array positioning and orientation. In \cite{ref13}, the authors investigated the Shannon information capacity and degrees of freedom in space-time wireless channels from the perspective of Maxwell's equations, incorporating radiated power and $L_2 $-norm constraints across various system models, including both near- and far-field scenarios. Separately, another study \cite{ref14} examined the channel capacity between two CAPAs using Kolmogorov information theory to quantify the maximum information that can be reliably transmitted over a wireless channel. Building on these efforts, the authors in \cite{ref15} introduced a rigorous analytical framework for evaluating mutual information and channel capacity in random fields, extending the analysis to non-white noise fields. Moreover, \cite{ref16} investigated CAPA systems in multipath fading channels, focusing on key performance metrics such as average data rate and outage probability, while also assessing the system's ergodic channel capacity. These works demonstrated that CAPAs could effectively utilize spatial resources to achieve higher communication rates and improved interference mitigation capabilities. 

In parallel, some efforts have focused on downlink beamforming for CAPA systems \cite{ref17, ref18, ref19, ref20}. In \cite{ref17}, a framework was proposed to characterize downlink capacity, analyzing the optimal single-user maximal-ratio transmission (MRT) beamforming scheme and designing a multi-user optimal power allocation strategy based on dirty paper coding (DPC) through uplink-downlink duality.  The study in \cite{ref18} introduced a wavenumber-division multiplexing (WDM) scheme, inspired by orthogonal frequency-division multiplexing (OFDM), to represent transmit currents and received fields using Fourier basis functions. However, this heuristic approach cannot achieve the optimal performance of CAPA systems. To address this, \cite{ref19} formulated the continuous current pattern design as a weighted sum-rate maximization problem. It proposed a Fourier-based discretization method, enabling the use of conventional optimization algorithms from discrete MIMO systems. This method has been applied to CAPA-based uplink beamforming \cite{ref21} and ISAC systems \cite{ref22}, and serves as a benchmark for the algorithm proposed in this paper. However, approximating continuous current functions with Fourier discretization leads to performance degradation. Increasing the number of Fourier bases can reduce this loss but increases computational complexity. To overcome this, \cite{ref20} proposed a calculus of variations (CoV) method to directly solve the functional programming problem, achieving optimal performance and significantly reducing computational complexity.

In addition to the aforementioned research, another equally important research direction is physical layer security. Multi-antenna technology has been shown to effectively enhance the secrecy performance of communication systems\cite{ref30}. The study in \cite{ref23} analyzes the upper bound of secrecy capacity in a MIMO system.
In \cite{ref24}, the secrecy performance of communication systems is further improved by leveraging the additional DOFs provided by reconfigurable intelligent surfaces (RIS). Compared to discrete MIMO, the use of CAPA allows for higher DOFs, which helps to strengthen the received signal at LUTs while degrading the received signal quality at Eves. The author in \cite{ref31} is the first to investigate secure transmission in a CAPA system involving a single LUT and a single Eve, focusing on the optimal current distribution and demonstrating the secrecy performance gain over discrete MIMO systems. However, in scenarios with multiple LUTs and multiple Eves, the benefits provided by a CAPA system remain unclear, necessitating the development of novel analytical tools. To address this gap, the current research aims to develop advanced secure beamforming strategies specifically tailored to CAPA systems with multiple LUTs and Eves.
The main contributions of this paper can be summarized as follows:
\begin{itemize}
    \item First, the secrecy performance of the CAPA-based communication system is represented by weighted secure sum-rate (WSSR), considering a downlink scenario with multiple LUTs and multiple Eves. Next, based on EM theory, the secure beamforming problem is modeled as a non-convex functional programming problem, aiming to explore the optimal design of the continuous source current patterns.
    
    \item We propose a low-complexity FP-based BCD optimization method to precisely optimize the continuous source current patterns without relying on any approximations, thereby obtaining the optimal solution to the secure beamforming problem. In this process, a continuous function inversion theory is introduced, providing a closed-form expression for the optimal current patterns.
    
    \item We further present a ZF-based heuristic secure beamforming approach for CAPA systems. This method completely eliminates interference among LUTs and prevents information leakage to Eves. Additionally, by incorporating a water-filling algorithm, it can obtain the optimal power allocation scheme for each LUT. The heuristic approach offers lower computational complexity.
    
    \item Extensive numerical results are presented to demonstrate the effectiveness of the proposed methods. The findings indicate that CAPA-based communication systems exhibit significantly superior secrecy performance compared to discrete MIMO systems. Additionally, the proposed FP-based BCD optimization method outperforms state-of-the-art Fourier-based techniques by achieving optimal WSSR performance across various system configurations, while drastically reducing computational complexity, particularly for large apertures and high frequencies. The heuristic ZF-based approach gradually approaches FP-based BCD optimization method at high SNR, and also outperforms the Fourier discretization method. Moreover, it delivers performance comparable to, or even exceeding, that of discrete MIMO optimization algorithms in certain scenarios.
\end{itemize}

The rest of this paper is structured as follows. Section II presents the CAPA-based secure transmission system and formulates the problem of WSSR maximization. Sections III and IV propose the FP-based BCD optimization approach and the heuristic ZF-based scheme for secure beamforming, respectively. Section V provides numerical results to compare the performance of various approaches across different system configurations. Finally, Section VI concludes the paper.

\section{System Model}
\begin{figure}[t!]
    \centering
    \includegraphics[width=0.3\textwidth]{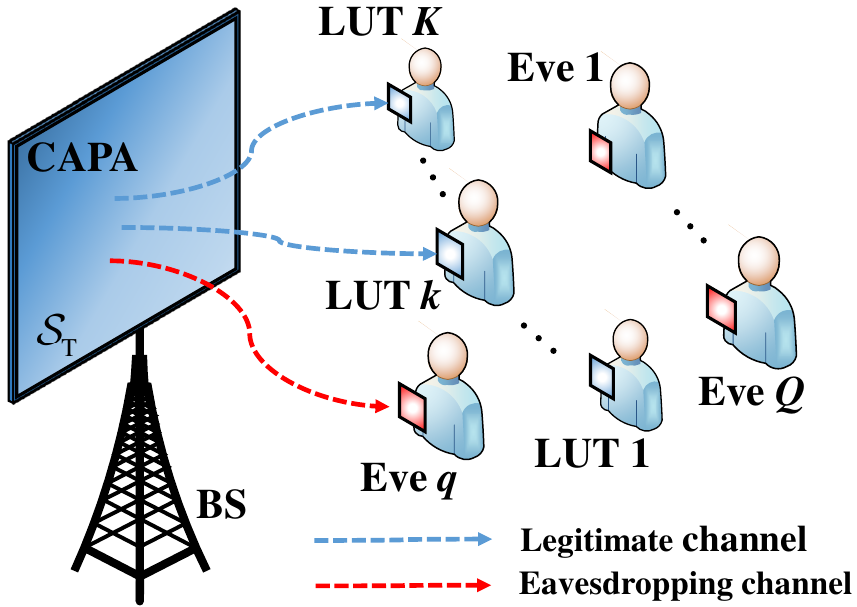} 
    \caption{Illustration of a CAPA-based downlink secure communication.}
    \label{fig:capa_system} 
\vspace{-15pt}
\end{figure}

A CAPA-based downlink secure transmission system is considered. We study a single cell where a base station (BS) with a CAPA serves $\mathit{K}$ single-antenna legitimate user terminals (LUTs). At the same time, there are $\mathit{Q}$ single-antenna eavesdroppers (Eves) overhearing the channel in the system. The CAPA-based BS is equipped with a continuous surface $\mathcal{S}_{\mathrm{T}}$, which has an area of $A_{\mathrm{T}}=\left|\mathcal{S}_{\mathrm{T}}\right|$. The schematic diagram is shown in Fig. \ref{fig:capa_system}. In a time division duplexing (TDD) system model, it is assumed that the Eves are disguised as legitimate registered users and can also send uplink pilots. Consequently, by utilizing the uplink-downlink reciprocity, the BS can obtain the Eves' channel state information (CSI) and use it for downlink secure beamforming, thereby protecting the transmitted confidential signals and minimizing information leakage.

\subsection{EM Signal Model}
For sake of brevity, we focus on a narrowband single-carrier communication system, where the source current distribution on the CAPA surface is independent of the current frequency and only related to the position of the surface. Then, the source current density distribution at point $\mathbf{s} \in \mathcal{S}_{\mathrm{T}}$ can be expressed as $\mathbf{J}(\mathbf{s})$.
In this paper, we consider a practically feasible vertically polarized transmitter. In this case, the current $\mathbf{J}(\mathbf{s})$ can also be represented as the product of a scalar current distribution and the polarization direction vector, i.e., $\mathbf{J}(\mathbf{s}) = J(\mathbf{s}) \hat{\mathbf{u}}_y$, where $\hat{\mathbf{u}}_y$ is a unit vector along the $y$-axis.

The signal samples transmitted from the BS to $K$ LUTs are denoted as $\mathbf{x}=[x_1, x_2, \ldots, x_K]^T\in \mathbb{C}^{K \times 1}$, which are assumed to be independent with unit power, satisfying $\mathbb{E}\left\{\mathbf{x} \mathbf{x}^H\right\}=\mathbf{I}_K$. Consequently, the information-carrying source current $J(\mathbf{s})$ can be represented as $J(\mathbf{s})=\sum_{k=1}^K J_k(\mathbf{s}) x_k$, where $J_k(\mathbf{s}) \in \mathbb{C}$ represents the source current density corresponding to the $k$-th LUT, analogous to the precoding vector in conventional MIMO systems.
Let $\mathbf{r}_k \in \mathbb{C}^{3\times 1}$ denote the location of the $k$-th LUT. The electric field at $\mathbf{r}_k$  excited by the source current $\mathbf{J}(\mathbf{s})$ can be expressed using the Green's function as \cite{ref6} 
\begin{align}
\mathbf{E}_k=\int_{S_{\mathrm{T}}} \mathbf{G}\left(\mathbf{r}_k, \mathbf{s}\right) \mathbf{J}(\mathbf{s}) \rm{d} \mathbf{s},
\label{electric_field}
\end{align}
where $\mathbf{E}_k \in \mathbb{C}^{3 \times 1}$ represents the electric field vector, and $\mathbf{G}\left(\mathbf{r}_k, \mathbf{s}\right)$ is the Green's function, given by
\begin{align}
\mathbf{G}(\mathbf{r}, \mathbf{s})=\frac{j \eta e^{-j \frac{2 \pi}{\lambda}\|\mathbf{r}-\mathbf{s}\|}}{2 \lambda\|\mathbf{r}-\mathbf{s}\|}\left(\mathbf{I}_3-\frac{(\mathbf{r}-\mathbf{s})(\mathbf{r}-\mathbf{s})^T}{\|\mathbf{r}-\mathbf{s}\|^2}\right),
\end{align}
where $\eta$ denotes the intrinsic impedance of the medium and $\lambda$ is the wavelength.

We assume that each LUT $k$ is equipped with a uni-polarized antenna oriented along the polarization direction $\hat{\mathbf{u}}_k$, where $\|\hat{\mathbf{u}}_k\|^2=1$. The antenna has an area of $A_{\mathrm{R}}=\frac{\lambda^2}{4\pi}$, with $A_{\mathrm{R}}\ll A_{\mathrm{T}}$. Given its negligible size, the antenna is approximated as a point in 3D space. Consequently, the noisy electric field observed by the receiver at LUT $k$ is
\begin{align}
&E_k=\hat{\mathbf{u}}_k^T \mathbf{E}_k+n_k =\int_{\mathcal{S}_{\mathrm{T}}} \hat{\mathbf{u}}_k^T \mathbf{G}\left(\mathbf{r}_k, \mathbf{s}\right) \mathbf{J}(\mathbf{s}) \rm{d} \mathbf{s}+n_k \\
&=\underbrace{\int_{S_{\mathrm{T}}} \hat{\mathbf{u}}_k^T \mathbf{G}\left(\mathbf{r}_k, \mathbf{s}\right) J_k(\mathbf{s}) x_k \hat{\mathbf{u}}_y \rm{d} \mathbf{s}}_{\text {desired signal }}\nonumber\\
&+\underbrace{\sum_{i=1, i \neq k}^K \int_{S_{\mathrm{T}}} \hat{\mathbf{u}}_k^T \mathbf{G}\left(\mathbf{r}_k, \mathbf{s}\right) J_i(\mathbf{s}) x_i \hat{\mathbf{u}}_y \rm{d} \mathbf{s}}_{\text {inter-LUT interference }}+n_k \\
&= \underbrace{\int_{S_{\mathrm{T}}} H_k(\mathbf{s}) J_k(\mathbf{s}) x_k \rm{d}\mathbf{s}}_{\text {desired signal }}+\underbrace{\sum_{i=1, i \neq k}^K \int_{S_{\mathrm{T}}} H_k(\mathbf{s}) J_i(\mathbf{s}) x_i \rm{d} \mathbf{s}}_{\text {inter-LUT interference }}+n_k,
\end{align}
Here, $H_k(\mathbf{s})=\hat{\mathbf{u}}_k^T \mathbf{G}\left(\mathbf{r}_k, \mathbf{s}\right) \hat{\mathbf{u}}_y \in \mathbb{C}$ represents the continuous EM legitimate channel to LUT $k$, and $n_k \in \mathbb{C}$ denotes the EM noise at LUT $k$, which can be modeled as additive white Gaussian noise (AWGN) with zero mean and variance $\sigma_k^2$, i.e., $n_k \sim \mathcal{C N}\left(0, \sigma_k^2\right)$ \cite{ref19}.

Similarly, the noisy electric field received by eavesdropper $q$ is written as
\begin{align}
\bar{E}_q&=\hat{\mathbf{u}}_q^T \bar{\mathbf{E}}_q+\bar{n}_{q}=\int_{\mathcal{S}_{\mathrm{T}}} \hat{\mathbf{u}}_q^T \mathbf{G}\left(\mathbf{r}_q, \mathbf{s}\right) \mathbf{J}(\mathbf{s}) \rm{d} \mathbf{s}+\bar{n}_q \\
& =\underbrace{\sum_{i=1}^K \int_{\mathcal{S}_{\mathrm{T}}} \hat{\mathbf{u}}_q^T \mathbf{G}\left(\mathbf{r}_q, \mathbf{s}\right) J_i(\mathbf{s}) x_i \hat{\mathbf{u}}_y \rm{d} \mathbf{s}}_{\text {all are desired signals }}+\bar{n}_q \\
&= \underbrace{\sum_{i=1}^K \int_{\mathcal{S}_{\mathrm{T}}} \bar{H}_q(\mathbf{s}) J_i(\mathbf{s}) x_i \rm{d} \mathbf{s}}_{\text {all are desired signals }}+\bar{n}_q
\end{align}
where $\bar{H}_q(\mathbf{s})=\hat{\mathbf{u}}_q^T \mathbf{G}\left(\mathbf{r}_q, \mathbf{s}\right) \hat{\mathbf{u}}_y\in \mathbb{C}$ represents the continuous EM eavesdropping channel to Eve $q$, and $ \bar{n}_q \sim \mathcal{C N}\left(0, \bar{\sigma}_q^2\right)\in \mathbb{C}$.

For LUT $k$, the maximum achievable rate during the downlink data transmission phase can be expressed as
\begin{align}
R_k = \log_2\left(1 + \gamma_k\right),
\end{align}
where $\gamma_k$ is the signal-to-interference-plus-noise ratio (SINR) at LUT $k$ given by
\begin{align}
\gamma_k=\frac{\left|\int_{\mathcal{S}_{\mathrm{T}}} H_k(\mathbf{s}) J_k(\mathbf{s}) \rm{d} \mathbf{s}\right|^2}{\sum_{i=1, i \neq k}^K\left|\int_{\mathcal{S}_{\mathrm{T}}} H_k(\mathbf{s}) J_i(\mathbf{s}) \rm{d} \mathbf{s}\right|^2+\sigma_k^2}.
\end{align}

To evaluate the performance of secure information transmission, we first introduce the secrecy rate $R_k^{\mathrm{sec}}$ for LUT $k$. It is defined as the downlink maximum achievable rate of LUT $k$ minus the rate of information leaked to all Eves. To simplify the expression of information leakage, we consider the worst-case scenario, assuming that all Eves can collaborate and completely eliminate the interference from all LUT data. Under this assumption, the information leakage rate can be expressed as:
\begin{align}
\bar{R}_k = \log_2\left(1 + \Gamma_{k}\right),
\end{align}
where $\Gamma_k$ is the aggregated signal-to-noise ratio (SNR) at all Eves given by
\begin{align}
\Gamma_{k}=\sum_{q=1}^Q\frac{\left|\int_{\mathcal{S}_{\mathrm{T}}} \bar{H}_q(\mathbf{s}) J_k(\mathbf{s}) \rm{d} \mathbf{s}\right|^2}{\bar{\sigma}_q^2}.
\end{align}
Thus, the secrecy rate of LUT $k$ can be expressed as $R_k^{\mathrm{sec}}=\left(R_k-\bar{R}_k\right)^{+}$, where $\left( x\right)^+=\mathrm{max}\{x,0\}$. The WSSR of the system is then given by $R^{\mathrm{sec}}=\sum_{k=1}^K \alpha_k R_k^{\mathrm{sec}}=\sum_{k=1}^K \alpha_k\left(\log _2\left(\frac{1+\gamma_k}{1+\Gamma_{k}}\right)\right)^{+}$, where $\{\alpha_k\}_{k=1}^{K}$ represents the non-negative priority weight corresponding to the quality of services (QoSs) of LUT $k$.

\subsection{Problem Formulation}
Our objective is to design the source current pattern $\{ J_k\left( \mathbf{s}\right)\}_{k=1}^K$ to maximize the WSSR. This leads to the following optimization problem:
\begin{subequations}\label{current_original}
\begin{align}
\max_{J_k(\mathbf{s})}~&R^{\mathrm{sec}}
\label{current_original_objective_function}\\
{\rm{s.t.}}~&\sum_{k=1}^{K}\int_{\mathcal{S}_{\mathrm{T}}}\left\lvert J_{k}(\mathbf{s})\right\rvert^2{\rm{d}}{\mathbf{s}}\leq P \label{power constraint},
\end{align}
\end{subequations}
where \eqref{power constraint} represents the maximum power constraint of the BS \cite{ref19}, and $P$ is the power budget.

To streamline notation, we normalize each spatial response by the corresponding noise power. Upon setting $H_k({\mathbf{s}})=\frac{H_k({\mathbf{s}})}{\sigma_k}$ and $\bar{H}_q({\mathbf{s}})=\frac{\bar{H}_q({\mathbf{s}})}{\bar{\sigma}_q}$, we rewrite \eqref{current_original} as \eqref{current_transformation0}, as shown at the bottom of the page.
\begin{figure*}[hb] 
\centering 
\hrule 
\vspace{-5pt} 
\begin{subequations}\label{current_transformation0}
\begin{align}
\max_{J_k(\mathbf{s})}~&\sum_{k=1}^{K}\alpha_k\left(\log_2\left(1+\frac{\left\lvert\int_{\mathcal{S}_{\mathrm{T}}}H_k({\mathbf{s}})J_k(\mathbf{s}){\rm{d}}{\mathbf{s}}\right\rvert^2}
{1+\sum_{i=1,i\ne k}^K\left\lvert\int_{\mathcal{S}_{\mathrm{T}}}H_k({\mathbf{s}})J_{i}(\mathbf{s}){\rm{d}}{\mathbf{s}}\right\rvert^2}\right)
-\log_2\left(1+\sum_{q=1}^{Q}\left\lvert\int_{\mathcal{S}_{\mathrm{T}}}\bar{H}_q({\mathbf{s}})J_{k}(\mathbf{s}){\rm{d}}{\mathbf{s}}\right\rvert^2\right)\right)^+\label{current_transformation0_Objective}\\
{\rm{s.t.}}~&\sum_{k=1}^{K}\int_{\mathcal{S}_{\mathrm{T}}}\left\lvert J_{k}(\mathbf{s})\right\rvert^2{\rm{d}}{\mathbf{s}}\leq P.
\end{align}
\end{subequations}
\end{figure*}
The problem \eqref{current_transformation0} involves the non-differentiable operator $(\cdot)^+$ and a non-convex objective function, making it challenging to solve. To tackle this optimization problem, two methods are proposed below.

\section{FP Based BCD Optimization Approach}
\subsection{Problem Reformulation}
The operator $(\cdot)^+$ in \eqref{current_transformation0_Objective} makes the objective of problem \eqref{current_transformation0} intractable. To take off this operator, we convert problem \eqref{current_transformation0} into its equivalent form as follows.
\vspace{-5pt}
\begin{lemma}\label{Lemma_Initial_Trans}
Given the spatial responses $\{H_k({\mathbf{s}})\}_{k=1}^{K}$ and $\{\bar{H}_q({\mathbf{s}})\}_{q=1}^{Q}$, problem \eqref{current_transformation0} is equivalent to the problem defined as follows:
\begin{subequations}\label{current_transformation1}
\begin{align}
\max_{J_k(\mathbf{s}),b_k}~&\sum_{k=1}^{K}b_k(\log_2\left(1+\gamma_k\right)
-\log_2\left(1+\Gamma_k\right))\label{current_transformation1_Objective}\\
{\rm{s.t.}}~&\sum_{k=1}^{K}\int_{\mathcal{S}_{\mathrm{T}}}\left\lvert J_{k}(\mathbf{s})\right\rvert^2{\rm{d}}{\mathbf{s}}\leq P,\\
&b_k\in[0,\alpha_k], \forall k.
\end{align}
\end{subequations}
\end{lemma}
\vspace{-5pt}

\begin{IEEEproof}
We prove this lemma by showing that \eqref{current_transformation1} can be equivalently transformed to \eqref{current_transformation0}. Given $\{J_k({\mathbf{s}})\}_{k=1}^{K}$, the optimal $b_k$ satisfies 
\begin{align}
b_k^{\star}=\begin{cases}
\alpha_k& \gamma_k\geq \Gamma_k,\\
0& \gamma_k< \Gamma_k.
\end{cases}.
\end{align}
Substituting $b_k=b_k^{\star}$ into \eqref{current_transformation1_Objective} gives
\begin{align}
\eqref{current_transformation1_Objective}
&=\sum_{k=1}^{K}b_k^{\star}(\log_2\left(1+\gamma_k\right)
-\log_2\left(1+\Gamma_k\right))\\
&=\sum_{k=1}^{K}\alpha_k\left(\log_2\left(1+\gamma_k\right)
-\log_2\left(1+\Gamma_k\right)\right)^+=\eqref{current_transformation0_Objective},
\end{align}
which establishes the equivalence between \eqref{current_transformation0} and \eqref{current_transformation1}.
\end{IEEEproof}

We further rewrite \eqref{current_transformation1_Objective} as follows:
\begin{align}\label{current_transformation1_Objective_transformation1}
\eqref{current_transformation1_Objective}=\sum_{k=1}^{K}b_k\left(\log_2\left(1+\gamma_k\right)
+\log_2\left(\frac{1}{1+\Gamma_k}\right)\right).
\end{align}
This \emph{sum-of-functions-of-ratio} form motivates us to exploit the fractional programming (FP) framework \cite{ref25} to further simplify \eqref{current_transformation1_Objective_transformation1}. We start the derivations by the following lemma:
\vspace{-5pt}
\begin{lemma}\label{Lemma_FP_Must}
Given the spatial responses $\{H_k({\mathbf{s}})\}_{k=1}^{K}$ and $\{\bar{H}_q({\mathbf{s}})\}_{q=1}^{Q}$, problem \eqref{current_transformation1} is equivalent to the problem defined as follows:
\begin{subequations}\label{current_transformation2}
\begin{align}
\max_{J_k(\mathbf{s}),b_k}~&\sum_{k=1}^{K}b_k\left(\log_2\left(1+\gamma_k\right)
+\log_2\left(1+\frac{G_{\Gamma}-\Gamma_k}{1+\Gamma_k}\right)\right)\label{current_transformation2_Objective}\\
{\rm{s.t.}}~&\sum_{k=1}^{K}\int_{\mathcal{S}_{\mathrm{T}}}\left\lvert J_{k}(\mathbf{s})\right\rvert^2{\rm{d}}{\mathbf{s}}\leq P,\\
&b_k\in[0,\alpha_k],\forall k,
\end{align}
\end{subequations}
where $G_{\Gamma}\triangleq P\sum_{q=1}^{Q}\int_{\mathcal{S}_{\mathrm{T}}}\left\lvert \bar{H}_q({\mathbf{s}})\right\rvert^2{\rm{d}}{\mathbf{s}}\geq \Gamma_k$.
\end{lemma}
\vspace{-5pt}
\begin{IEEEproof}
According to the Cauchy–Schwarz inequality, we have
\begin{subequations}
\begin{align}
\Gamma_{k}&=\sum_{q=1}^{Q}{\left\lvert\int_{\mathcal{S}_{\mathrm{T}}}\bar{H}_q({\mathbf{s}})J_{k}(\mathbf{s}){\rm{d}}{\mathbf{s}}\right\rvert^2}\nonumber\\
&\leq\sum_{q=1}^{Q}\int_{\mathcal{S}_{\mathrm{T}}}\left\lvert \bar{H}_q({\mathbf{s}})\right\rvert^2{\rm{d}}{\mathbf{s}}\int_{\mathcal{S}_{\mathrm{T}}}\left\lvert J_{k}(\mathbf{s})\right\rvert^2{\rm{d}}{\mathbf{s}}\\
&\leq P\sum_{q=1}^{Q}\int_{\mathcal{S}_{\mathrm{T}}}\left\lvert \bar{H}_q({\mathbf{s}})\right\rvert^2{\rm{d}}{\mathbf{s}}=G_{\Gamma}.
\end{align}
\end{subequations}
It follows that
\begin{subequations}
\begin{align}
\log_2\left(\frac{1}{1+\Gamma_{k}}\right)&=\log_2\left(\frac{1+G_{\Gamma}}{1+\Gamma_{k}}\right)-\log_2\left(1+G_{\Gamma}\right)\\
&=\log_2\left(1+\frac{G_{\Gamma}-\Gamma_{k}}{1+\Gamma_{k}}\right)-\log_2\left(1+G_{\Gamma}\right).
\end{align}
\end{subequations}
Note that $\log_2\left(1+G_{\Gamma}\right)$ is independent of $\{J_k({\mathbf{s}})\}_{k=1}^{K}$ and $\{b_k\}_{k=1}^{K}$. Hence, we can establish the equivalence between \eqref{current_transformation1} and \eqref{current_transformation2}.
\end{IEEEproof}

We next exploit the FP framework to convert the fractional programming problem \eqref{current_transformation2} equivalently to a new problem which involves a more tractable objective function.
\vspace{-5pt}
\begin{lemma}\label{Lemma_Dual}
Given the spatial responses $\{H_k({\mathbf{s}})\}_{k=1}^{K}$ and $\{\bar{H}_q({\mathbf{s}})\}_{q=1}^{Q}$, problem \eqref{current_transformation2} is equivalent to the problem defined as \eqref{current_transformation3} at the bottom of this page.
\begin{figure*}[hb] 
\centering 
\hrule 
\begin{subequations}\label{current_transformation3}
\begin{align}
\max_{J_k(\mathbf{s}),b_k,\epsilon_k,\beta_k}~&\sum_{k=1}^{K}b_k\left(\log(1+\epsilon_k)-\epsilon_k+
\frac{(1+\epsilon_k)\left\lvert\int_{\mathcal{S}_{\mathrm{T}}}H_k({\mathbf{s}})J_k(\mathbf{s}){\rm{d}}{\mathbf{s}}\right\rvert^2}
{1+\sum_{i=1}^{K}\left\lvert\int_{\mathcal{S}_{\mathrm{T}}}H_k({\mathbf{s}})J_{i}(\mathbf{s}){\rm{d}}{\mathbf{s}}\right\rvert^2}+
\log(1+\beta_k)-\beta_k+(1+\beta_k)\frac{G_{\Gamma}-\Gamma_k}{1+G_{\Gamma}}\right)\label{current_transformation3_Objective}\\
{\rm{s.t.}}~&\sum_{k=1}^{K}\int_{\mathcal{S}_{\mathrm{T}}}\left\lvert J_{k}(\mathbf{s})\right\rvert^2{\rm{d}}{\mathbf{s}}\leq P,\\
&b_k\in[0,\alpha_k], \forall k.
\end{align}
\end{subequations}
\end{figure*}
The optimal $\epsilon_k$ and $\beta_k$ are given by $\epsilon_k^{\star}=\gamma_k$ and $\beta_k^{\star}=\frac{G_{\Gamma}-\Gamma_k}{1+\Gamma_k}$, respectively.
\end{lemma}
\vspace{-5pt}
\begin{IEEEproof}
Since \eqref{current_transformation3_Objective} is concave over $\epsilon_k$ and $\beta_k$ for fixed $\{J_k({\mathbf{s}})\}_{k=1}^{K}$ and $\{b_k\}_{k=1}^{K}$, we take its complex derivative and solve each 
\begin{subequations}
\begin{align}
\frac{\partial}{\partial \epsilon_k}\eqref{current_transformation3_Objective}&=\frac{1}{1+\epsilon_k}-1+\frac{\left\lvert\int_{\mathcal{S}_{\mathrm{T}}}H_k({\mathbf{s}})J_k(\mathbf{s}){\rm{d}}{\mathbf{s}}\right\rvert^2}
{1+\sum_{i=1}^{K}\left\lvert\int_{\mathcal{S}_{\mathrm{T}}}H_k({\mathbf{s}})J_{i}(\mathbf{s}){\rm{d}}{\mathbf{s}}\right\rvert^2}=0,\\
\frac{\partial}{\partial \beta_k}\eqref{current_transformation3_Objective}&=\frac{1}{1+\beta_k}-1+\frac{G_{\Gamma}-\Gamma_k}{1+G_{\Gamma}}=0.
\end{align}
\end{subequations}
The optimal $\epsilon_k$ and $\beta_k$ are easily seen as 
\begin{subequations}
\begin{align}
\epsilon_k^{\star}&=\frac{\left\lvert\int_{\mathcal{S}_{\mathrm{T}}}H_k({\mathbf{s}})J_k(\mathbf{s}){\rm{d}}{\mathbf{s}}\right\rvert^2}
{1+\sum_{i=1,i\ne k}^{K}\left\lvert\int_{\mathcal{S}_{\mathrm{T}}}H_k({\mathbf{s}})J_{i}(\mathbf{s}){\rm{d}}{\mathbf{s}}\right\rvert^2}=\gamma_k,\\
\beta_k^{\star}&=\frac{G_{\Gamma}-\Gamma_k}{1+\Gamma_k},
\end{align}
\end{subequations}
respectively. Inserting $\alpha_k^{\star}$ and $\beta_k^{\star}$ back to \eqref{current_transformation3_Objective} recovers the objective function in \eqref{current_transformation2_Objective}, thus establishing the equivalence of these two problems.
\end{IEEEproof}
It is worth noting that $G_{\Gamma}$ is independent of $\{J_k({\mathbf{s}}),b_k,\epsilon_k,\beta_k\}_{k=1}^{K}$. Therefore, the intractability of \eqref{current_transformation3_Objective} mainly originates from the fractional terms involved in $\left\{\frac{(1+\epsilon_k)\left\lvert\int_{\mathcal{S}_{\mathrm{T}}}H_k({\mathbf{s}})J_k(\mathbf{s}){\rm{d}}{\mathbf{s}}\right\rvert^2}{1+\sum_{i=1}^{K}\left\lvert\int_{\mathcal{S}_{\mathrm{T}}}H_k({\mathbf{s}})J_{i}(\mathbf{s}){\rm{d}}{\mathbf{s}}\right\rvert^2}\right\}_{k=1}^{K}$. To handle this difficulty, we introduce auxiliary variables $\{\eta_k\}_{k=1}^{K}$ to simplify \eqref{current_transformation3_Objective}\cite{ref26}. Then, the following lemma can be found.
\vspace{-5pt}
\begin{lemma}\label{Lemma_Quad}
Given the spatial responses $\{H_k({\mathbf{s}})\}_{k=1}^{K}$ and $\{\bar{H}_q({\mathbf{s}})\}_{q=1}^{Q}$, problem \eqref{current_transformation3} is equivalent to the problem defined as \eqref{current_transformation4} at the bottom of next page.
\begin{figure*}[hb] 
\centering 
\hrule 
\begin{subequations}\label{current_transformation4}
\begin{align}
\max_{J_k(\mathbf{s}),b_k,\epsilon_k,\beta_k, \eta_k}~&\sum_{k=1}^{K}b_k\left(\log(1+\epsilon_k)-\epsilon_k+(1+\epsilon_k)\left(2\Re\left\{{\eta}_k^{*}
\int_{\mathcal{S}_\mathrm{T}}H_k({\mathbf{s}})J_k(\mathbf{s}){\rm{d}}{\mathbf{s}}\right\}
-\left\lvert\eta_k\right\rvert^2\left(1+\sum_{i=1}^{K}\left\lvert\int_{\mathcal{S}_\mathrm{T}}H_k({\mathbf{s}})J_{i}(\mathbf{s}){\rm{d}}{\mathbf{s}}\right\rvert^2
\right)\right)\right.\nonumber\\
&+\left.\log(1+\beta_k)-\beta_k+(1+\beta_k)\frac{G_{\Gamma}-\Gamma_k}{1+G_{\Gamma}}\right)\label{current_transformation4_Objective}\\
{\rm{s.t.}}~&\sum_{k=1}^{K}\int_{\mathcal{S}_\mathrm{T}}\left\lvert J_{k}(\mathbf{s})\right\rvert^2{\rm{d}}{\mathbf{s}}\leq P,\\
&b_k\in[0,\alpha_k], \forall k.
\end{align}
\end{subequations}
\end{figure*}
The optimal $\eta_k$ is given by $\eta_k^{\star}=\left(1+\sum_{i=1}^{K}\left\lvert\int_{\mathcal{S}_\mathrm{T}}H_k({\mathbf{s}})J_{i}(\mathbf{s}){\rm{d}}{\mathbf{s}}\right\rvert^2
\right)^{-1}\int_{\mathcal{S}_\mathrm{T}}H_k({\mathbf{s}})J_k(\mathbf{s}){\rm{d}}{\mathbf{s}}$.
\end{lemma}
\vspace{-5pt}
\begin{IEEEproof}
Since \eqref{current_transformation4_Objective} is concave over $\eta_k$ for fixed $\{J_k({\mathbf{s}}),b_k,\epsilon_k,\beta_k\}_{k=1}^{K}$, we take its complex derivative and solve  
\begin{align}
\frac{\partial}{\partial \eta_k}\eqref{current_transformation4_Objective}&=
\int_{\mathcal{S}_\mathrm{T}}H_k({\mathbf{s}})J_k(\mathbf{s}){\rm{d}}{\mathbf{s}}\nonumber\\
&-\eta_k\left(1+\sum_{i=1}^{K}\left\lvert\int_{\mathcal{S}_\mathrm{T}}H_k({\mathbf{s}})J_{i}(\mathbf{s}){\rm{d}}{\mathbf{s}}\right\rvert^2
\right)=0.
\end{align}
The optimal $\eta_k$ are easily seen as 
\begin{align}
\eta_k^{\star}=\frac{\int_{\mathcal{S}_\mathrm{T}}H_k({\mathbf{s}})J_k(\mathbf{s}){\rm{d}}{\mathbf{s}}}
{1+\sum_{i=1}^{K}\left\lvert\int_{\mathcal{S}_\mathrm{T}}H_k({\mathbf{s}})J_{i}(\mathbf{s}){\rm{d}}{\mathbf{s}}\right\rvert^2}.
\end{align}
Inserting $\eta_k^{\star}$ back to \eqref{current_transformation4_Objective} recovers the objective function in \eqref{current_transformation3_Objective}, thus establishing the equivalence of these two problems.
\end{IEEEproof}
\subsection{FP-based BCD Algorithm}
At this point, the transformed problem \eqref{current_transformation4_Objective} is a joint optimization problem involving $\{J_k({\mathbf{s}}),b_k,\epsilon_k,\beta_k,\eta_k\}_{k=1}^{K}$, 
where the variables are decoupled from each other, and the BCD algorithm can be used for alternating optimization.

\subsubsection{Optimization of $b_k$}
By fixing $\{J_k({\mathbf{s}})\}_{k=1}^K$, the marginal problem for $b_k$ is given by
\begin{subequations}
\begin{align}
\max_{b_k}~&b_k(\log_2\left(1+\gamma_k\right)
-\log_2\left(1+\Gamma_k\right))\\
{\rm{s.t.}}~&b_k\in[0,\alpha_k].
\end{align}
\end{subequations}
According to \textbf{Lemma~\ref{Lemma_Initial_Trans}}, the solution can be obtained as:
\begin{align}\label{b_k}
b_k=\begin{cases}
\alpha_k& \gamma_k\geq \Gamma_k,\\
0& \gamma_k< \Gamma_k.
\end{cases}
\end{align}

\subsubsection{Optimization of $\epsilon_k$}
By fixing $\{J_k({\mathbf{s}}), b_k,\beta_k\}_{k=1}^K$, the marginal problem for $\epsilon_k$ is given by
\begin{align}
\max_{\epsilon_k}&
\left(
\log(1+\epsilon_k)-\epsilon_k+
\frac{(1+\epsilon_k)\left\lvert\int_{\mathcal{S}_{\mathrm{T}}}H_k({\mathbf{s}})J_k(\mathbf{s}){\rm{d}}{\mathbf{s}}\right\rvert^2}
{1+\sum_{i=1}^{K}\left\lvert\int_{\mathcal{S}_{\mathrm{T}}}H_k({\mathbf{s}})J_{i}(\mathbf{s}){\rm{d}}{\mathbf{s}}\right\rvert^2}
\right),
\end{align}
According to \textbf{Lemma~\ref{Lemma_Dual}}, the solution can be obtained as:
\begin{align}\label{epsilon_k}
\epsilon_k&=\frac{\left\lvert\int_{\mathcal{S}_{\mathrm{T}}}H_k({\mathbf{s}})J_k(\mathbf{s}){\rm{d}}{\mathbf{s}}\right\rvert^2}
{1+\sum_{i=1,i\ne k}^{K}\left\lvert\int_{\mathcal{S}_{\mathrm{T}}}H_k({\mathbf{s}})J_{i}(\mathbf{s}){\rm{d}}{\mathbf{s}}\right\rvert^2}=\gamma_k.
\end{align}

\subsubsection{Optimization of $\beta_k$}
By fixing $\{J_k({\mathbf{s}}), b_k,\epsilon_k\}_{k=1}^K$, the marginal problem for $\beta_k$ is given by
\begin{align}
\max_{\beta_k}&
\left(
\log(1+\beta_k)-\beta_k+(1+\beta_k)\frac{G_{\Gamma}-\Gamma_k}{1+G_{\Gamma}}
\right),
\end{align}
According to \textbf{Lemma~\ref{Lemma_Dual}}, the solution can be obtained as:
\begin{align}\label{beta_k}
\beta_k=\frac{G_{\Gamma}-\Gamma_k}{1+\Gamma_k}.
\end{align}

\subsubsection{Optimization of $\eta_k$}
By fixing $\{J_k({\mathbf{s}}), b_k,\beta_k,\epsilon_k\}_{k=1}^K$, and noting that $\epsilon_k\geq0$, the marginal problem for $\eta_k$ is given by
\begin{align}
\max_{\eta_k}~&\left(2\Re\left\{{\eta}_k^{*}\int_{\mathcal{S}_\mathrm{T}}H_k({\mathbf{s}})J_k(\mathbf{s}){\rm{d}}{\mathbf{s}}\right\}\right.\nonumber\\
&-\left.\left\lvert\eta_k\right\rvert^2\left(1+\sum_{i=1}^{K}\left\lvert\int_{\mathcal{S}_\mathrm{T}}H_k({\mathbf{s}})J_{i}(\mathbf{s}){\rm{d}}{\mathbf{s}}\right\rvert^2
\right)
\right),
\end{align}
According to \textbf{Lemma~\ref{Lemma_Quad}}, the solution can be obtained as:
\begin{align}\label{eta_k}
\eta_k=\frac{\int_{\mathcal{S}_\mathrm{T}}H_k({\mathbf{s}})J_k(\mathbf{s}){\rm{d}}{\mathbf{s}}}
{1+\sum_{i=1}^{K}\left\lvert\int_{\mathcal{S}_\mathrm{T}}H_k({\mathbf{s}})J_{i}(\mathbf{s}){\rm{d}}{\mathbf{s}}\right\rvert^2}.
\end{align}

\subsubsection{Optimization of $J_k(\mathbf{s})$}
By fixing $\{b_k,\beta_k,\epsilon_k, \eta_k\}_{k=1}^K$, the marginal problem for $J_k(\mathbf{s})$ is given by \eqref{only_current_optimize},
\begin{figure*}[!b] 
\centering 
\hrule 
\vspace{-1pt} 
\begin{subequations}\label{only_current_optimize}
\begin{align}
\min_{J_k(\mathbf{s})}~&\sum_{k=1}^{K}b_k\left((1+\epsilon_k)\left(\left\lvert\eta_k\right\rvert^2\sum_{i=1}^{K}\left\lvert\int_{\mathcal{S}_\mathrm{T}}H_k({\mathbf{s}})J_{i}(\mathbf{s}){\rm{d}}{\mathbf{s}}\right\rvert^2-
2\Re\left\{{\eta}_k^{*}\int_{\mathcal{S}_\mathrm{T}}H_k({\mathbf{s}})J_k(\mathbf{s}){\rm{d}}{\mathbf{s}}\right\}\right)+
\frac{\sum_{q=1}^{Q}\left\lvert\int_{\mathcal{S}_\mathrm{T}}\bar{H}_q({\mathbf{s}})J_{k}(\mathbf{s}){\rm{d}}{\mathbf{s}}\right\rvert^2}{1+G_{\Gamma}}(1+\beta_k)\right)\\
{\rm{s.t.}}~&\sum_{k=1}^{K}\int_{\mathcal{S}_\mathrm{T}}\left\lvert J_{k}(\mathbf{s})\right\rvert^2{\rm{d}}{\mathbf{s}}\leq P,
\end{align}
\end{subequations}
\hrule 
\end{figure*}
whose objective function can be rewritten as \eqref{only_current_optimize_trans}. Both \eqref{only_current_optimize} and \eqref{only_current_optimize_trans} are shown at the bottom of next page.
\begin{figure*}[!b] 
\centering 
\vspace{-10pt} 
\begin{equation}\label{only_current_optimize_trans}
\begin{split}
f(\{J_{k}(\mathbf{s})\}_{k=1}^K)&=\sum_{k=1}^{K}\left(\sum_{i=1}^{K}b_{i}(1+\epsilon_{i})\left\lvert\eta_{i}\right\rvert^2\left\lvert\int_{\mathcal{S}_\mathrm{T}}H_{i}({\mathbf{s}})J_{k}(\mathbf{s}){\rm{d}}{\mathbf{s}}\right\rvert^2
+\frac{1+\beta_k}{1+G_{\Gamma}}\sum_{q=1}^{Q}\left\lvert\int_{\mathcal{S}_\mathrm{T}}\bar{H}_q({\mathbf{s}})J_{k}(\mathbf{s}){\rm{d}}{\mathbf{s}}\right\rvert^2\right.\\
&-\left.2b_k(1+\epsilon_k)\Re\left\{{\eta}_k^{*}\int_{\mathcal{S}_\mathrm{T}}H_k({\mathbf{s}})J_k(\mathbf{s}){\rm{d}}{\mathbf{s}}\right\}\right)=\sum_{k=1}^{K}f(J_{k}(\mathbf{s})).
\end{split}
\end{equation}
\end{figure*}
Since the objective function is convex and the constraint set is convex, the Karush–Kuhn–Tucker (KKT) conditions are sufficient for optimality. The Lagrangian function is given by
\begin{subequations}
\begin{align}\label{Lagrange_Function_1}
{\mathcal{L}}=&\sum_{k=1}^{K}f(J_{k}(\mathbf{s}))+\lambda\left(\sum_{k=1}^{K}\int_{\mathcal{S}_\mathrm{T}}\left\lvert J_{k}(\mathbf{s})\right\rvert^2{\rm{d}}{\mathbf{s}}- P\right)\\
=&\sum_{k=1}^{K}g(J_{k}(\mathbf{s}))-\lambda P\label{Lagrange_Function_to_g},
\end{align}
\end{subequations}
where $\lambda\geq0$ is the Lagrange multiplier and $g(J_{k}(\mathbf{s}))$ can be expressed as
\begin{align}\label{Lagrange_Function}
g(J_{k}(\mathbf{s}))=f(J_{k}(\mathbf{s}))+\lambda\int_{\mathcal{S}_\mathrm{T}}\left\lvert J_{k}(\mathbf{s})\right\rvert^2{\rm{d}}{\mathbf{s}}.
\end{align}
From \eqref{Lagrange_Function_to_g}, it can be seen that finding the minimum of the Lagrangian function $\mathcal{L}$ is essentially equivalent to finding each $J_{k}(\mathbf{s})$ that minimizes the function $g(J_{k}(\mathbf{s}))$. To address this problem, we provide the following \textbf{Theorem~\ref{optimal_j_cov}}.
\begin{theorem}\label{optimal_j_cov}
The functional $g(J_k(\mathbf{s}))$ achieves its minimum when $J_k(\mathbf{s})$ takes the following form:
\begin{align}\label{optimal_J_structure}
b_k&(1+\epsilon_k)\eta_k H_k^*(\mathbf{s})= \lambda{J}_k(\mathbf{s})\nonumber\\
&+\sum_{i=1}^K b_i(1+\epsilon_i)|\eta_i|^2 H_i^*(\mathbf{s}) \int_{S_T} H_i(\mathbf{s}'){J}_k(\mathbf{s}') \, d\mathbf{s}' \nonumber \\
&+\frac{1+\beta_k}{1+G_{\Gamma}}\sum_{q=1}^Q  \bar{H}_q^*(\mathbf{s}) \int_{S_T} \bar{H}_q(\mathbf{s}'){J}_k(\mathbf{s}') \, d\mathbf{s}'.
\end{align}
\end{theorem}
\begin{IEEEproof}
Please refer to Appendix~\ref{Appendix_F}.
\end{IEEEproof}
Observing \eqref{optimal_J_structure}, it follows that
\begin{equation}\label{W_equal}
\int_{\mathcal{S}_\mathrm{T}}W_{\lambda,k}({\mathbf{s}},{\mathbf{s}}')J_{k}(\mathbf{s}'){\rm{d}}{\mathbf{s}}'=b_k(1+\epsilon_k){\eta}_kH_k^*({\mathbf{s}}),
\end{equation}
where
\begin{align}\label{W_original}
W_{\lambda,k}({\mathbf{s}},{\mathbf{s}}')=\lambda\delta({\mathbf{s}}-{\mathbf{s}}')&+\sum_{i=1}^{K}b_{i}(1+\epsilon_{i})\left\lvert\eta_{i}\right\rvert^2H_{i}^*({\mathbf{s}})H_{i}({\mathbf{s}}')\nonumber\\
&+\frac{1+\beta_k}{1+G_{\Gamma}}\sum_{q=1}^{Q}\bar{H}_q^*({\mathbf{s}})\bar{H}_q({\mathbf{s}}'),
\end{align}
and where $\delta(\cdot)$ denotes the Dirac delta function. 

The subsequent task is to find the inversion of $W_{\lambda,k}({\mathbf{s}},{\mathbf{s}}')$, denoted as ${\hat{W}}_{\lambda,k}({\mathbf{s}}'',{\mathbf{s}})$, which satisfies 
\begin{align}\label{Inversion_Condition}
\int_{{\mathcal{S}_\mathrm{T}}}&{\hat{W}}_{\lambda,k}({\mathbf{s}}'',{\mathbf{s}})W_{\lambda,k}({\mathbf{s}},{\mathbf{s}}'){\rm{d}}{\mathbf{s}}\nonumber\\
&=\int_{\mathcal{S}_\mathrm{T}}W_{\lambda,k}({\mathbf{s}}'',{\mathbf{s}}){\hat{W}}_{\lambda,k}({\mathbf{s}},{\mathbf{s}}'){\rm{d}}{\mathbf{s}}
=\delta({\mathbf{s}}''-{\mathbf{s}}').
\end{align}
Under this circumstance, we would have
\begin{align}\label{Inversion_Operation}
\int_{\mathcal{S}_\mathrm{T}}&{\hat{W}}_{\lambda,k}({\mathbf{s}}'',{\mathbf{s}})\int_{\mathcal{S}_\mathrm{T}} W_{\lambda,k}({\mathbf{s}},{\mathbf{s}}')J_{k}(\mathbf{s}'){\rm{d}}{\mathbf{s}}'{\rm{d}}{\mathbf{s}}\nonumber\\
&=\int_{\mathcal{S}_\mathrm{T}}\int_{\mathcal{S}_\mathrm{T}}{\hat{W}}_{\lambda,k}({\mathbf{s}}'',{\mathbf{s}})W_{\lambda,k}({\mathbf{s}},{\mathbf{s}}'){\rm{d}}{\mathbf{s}}J_{k}(\mathbf{s}'){\rm{d}}{\mathbf{s}}'
=J_{k}(\mathbf{s}'').
\end{align}
The main results are given in the following theorem.
\begin{theorem}\label{theorem1}
Given ${\Psi}({\mathbf{s}},{\mathbf{s}}')=\delta({\mathbf{s}}-{\mathbf{s}}')+\sum_{i=1}^{I}\psi_{i}^*({\mathbf{s}})\psi_{i}({\mathbf{s}}')$, its inversion is given by
\begin{align}\label{Inversion_Operator}
\hat{\Psi}({\mathbf{s}}'',{\mathbf{s}})=\delta({\mathbf{s}}''-{\mathbf{s}})-\sum_{i=1}^{I}\sum_{i'=1}^{I}\psi_{i}^*({\mathbf{s}}'')\psi_{i'}({\mathbf{s}})[({\mathbf{I}}+{\bm{\Psi}})^{-1}]_{i,i'},
\end{align}
where $[({\mathbf{I}}+{\bm{\Psi}})^{-1}]_{i,i'}$ is the $(i,i')$-th element of matrix $({\mathbf{I}}+{\bm{\Psi}})^{-1}$ with $[{\bm{\Psi}}]_{i,i'}=\int_{\mathcal{S}_{\mathrm{T}}}\psi_{i}({\mathbf{s}})\psi_{i'}^*({\mathbf{s}}){\rm{d}}{\mathbf{s}}$.
\end{theorem}
\begin{IEEEproof}
Please refer to Appendix~\ref{Appendix_E}.
\end{IEEEproof}

We next use the \textbf{Theorem~\ref{theorem1}} to calculate the inverse of $W_{\lambda,k}({\mathbf{s}},{\mathbf{s}}')$ in \eqref{W_original}, which is given by
\vspace{-8pt}
\begin{align}\label{W_inversion}
&{\hat{W}}_{\lambda,k}({\mathbf{s}}'',{\mathbf{s}})=\frac{1}{\lambda}\Bigg(\delta({\mathbf{s}}''-{\mathbf{s}})\nonumber\\
&\qquad-\sum_{n=1}^{N}\sum_{n'=1}^{N}\phi_{k,n}^*({\mathbf{s}}'')\phi_{k,n'}({\mathbf{s}})
[(\lambda{\mathbf{I}}+{\bm{\Phi}}_k)^{-1}]_{n,n'}\Bigg),
\end{align}
where
\vspace{-5pt}
\begin{subequations}\label{phi}
\begin{align}
\phi_{k,i}({\mathbf{s}})&=\sqrt{b_{i}(1+\epsilon_{i})\left\lvert\eta_{i}\right\rvert^2}H_{i}({\mathbf{s}}),~i=1,\ldots,K,\\
\phi_{k,K+q}({\mathbf{s}})&=\sqrt{\frac{1+\beta_k}{1+G_{\Gamma}}}\bar{H}_q({\mathbf{s}}),~q=1,\ldots,Q,\\
N&=K+Q,\\
[{\bm{\Phi}_k}]_{n,n'}&=\int_{\mathcal{S}_{\mathrm{T}}}\phi_{k,n}({\mathbf{s}})\phi_{k,n'}^*({\mathbf{s}}){\rm{d}}{\mathbf{s}}.
\end{align}
\end{subequations}
It follows from \eqref{W_equal} that
\begin{subequations}
\begin{align}
J_{k}(\mathbf{s}'')&=b_k(1+\epsilon_k){\eta}_k\int_{\mathcal{S}_{\mathrm{T}}}{\hat{W}}_{\lambda,k}({\mathbf{s}}'',{\mathbf{s}})H_k^*({\mathbf{s}}){\rm{d}}{\mathbf{s}}\\
&=\int_{\mathcal{S}_{\mathrm{T}}}{\hat{W}}_{\lambda,k}({\mathbf{s}}'',{\mathbf{s}})\frac{b_k(1+\epsilon_k){\eta}_k\phi_{k,k}^*({\mathbf{s}})}{\sqrt{b_{k}(1+\epsilon_{k})\left\lvert\eta_{k}\right\rvert^2}}{\rm{d}}{\mathbf{s}}\label{J_k}.
\end{align}
\end{subequations}
For brevity, we denote $a_{k,k}=\frac{b_k(1+\epsilon_k){\eta}_k}{\sqrt{b_{k}(1+\epsilon_{k})\left\lvert\eta_{k}\right\rvert^2}}=\frac{\sqrt{b_k(1+\epsilon_k)}{\eta}_k}{\left\lvert\eta_{k}\right\rvert}$. Substituting \eqref{W_inversion} into \eqref{J_k}, we obtain:
\begin{subequations}
\begin{align}
&J_{k}(\mathbf{s}'')=\frac{a_{k,k}}{\lambda}\Bigg(\phi_{k,k}^*(\mathbf{s}'')-\nonumber\\
&\sum_{n=1}^{N}\sum_{n'=1}^{N}\phi_{k,n}^*({\mathbf{s}}'')\int_{\mathcal{S}_{\mathrm{T}}}\phi_{k,n'}({\mathbf{s}})\phi_{k,k}^*({\mathbf{s}}){\rm{d}}{\mathbf{s}}
[(\lambda{\mathbf{I}}+{\bm{\Phi}}_k)^{-1}]_{n,n'}\Bigg)\\
&=\frac{a_{k,k}}{\lambda}\left(\phi_{k,k}^*(\mathbf{s}'')-
\sum_{n=1}^{N}\phi_{k,n}^*({\mathbf{s}}'')[(\lambda{\mathbf{I}}+{\bm{\Phi}}_k)^{-1}{\bm\phi}_{k,k}]_{n}\right),
\end{align}
\end{subequations}
where ${\bm\phi}_{k,k}$ denotes the $k$-th column of matrix ${\bm\Phi}_k$, and $[(\lambda{\mathbf{I}}+{\bm{\Phi}}_k)^{-1}{\bm\phi}_{k,k}]_{n}$ denotes the $n$-th element of $(\lambda{\mathbf{I}}+{\bm{\Phi}}_k)^{-1}{\bm\phi}_{k,k}$. 
To provide a more compact expression, we rewrite $J_{k}(\mathbf{s}'')$ as follows:
\begin{align}\label{J_k_compact}
J_{k}(\mathbf{s}'')=\frac{a_{k,k}}{\lambda}\sum_{n=1}^{N}[{\mathbf{w}}_k]_n\phi_{k,n}^*(\mathbf{s}''),
\end{align}
where the $n$-th element in vector ${\mathbf{w}}_k$ is given by
\begin{align}
[{\mathbf{w}}_k]_n=\begin{cases}
1-[(\lambda{\mathbf{I}}+{\bm{\Phi}}_k)^{-1}{\bm\phi}_{k,k}]_{n}& n=k,\\
-[(\lambda{\mathbf{I}}+{\bm{\Phi}}_k)^{-1}{\bm\phi}_{k,k}]_{n}& n\ne k.
\end{cases}
\end{align}

Furthermore, to determine the optimal value of $\lambda$, the transmit power of the LUT $k$ can be expressed as:
\begin{align}
&\int_{\mathcal{S}_{\mathrm{T}}}\left\lvert J_{k}(\mathbf{s}'')\right\rvert^2{\rm{d}}{\mathbf{s}''}\nonumber\\
&\quad=\int_{\mathcal{S}_{\mathrm{T}}}\left\lvert\frac{a_{k,k}}{\lambda}\right\rvert^2\sum_{n'=1}^{N}[{\mathbf{w}}_k]_{n'}\phi_{k,n'}^*(\mathbf{s}'')\sum_{n=1}^{N}[{\mathbf{w}}_k]_{n}^{*}\phi_{k,n}(\mathbf{s}''){\rm{d}}{\mathbf{s}''}\nonumber\\
&\quad=\left\lvert\frac{a_{k,k}}{\lambda}\right\rvert^2\sum_{n=1}^{N}\sum_{n'=1}^{N}[{\mathbf{w}}_k]_{n}^{*}\int_{\mathcal{S}_{\mathrm{T}}}\phi_{k,n}(\mathbf{s}'')\phi_{k,n'}^*(\mathbf{s}''){\rm{d}}{\mathbf{s}''}[{\mathbf{w}}_k]_{n'}\nonumber\\
&\quad=\left\lvert\frac{a_{k,k}}{\lambda}\right\rvert^2{\mathbf{w}}_k^{\mathsf{H}}{\bm\Phi}_k{\mathbf{w}}_k.
\end{align}
For clarity, we use ${\mathbf{U}}_k^{\mathsf{H}}{\bm\Lambda}_k{\mathbf{U}}_k$ denote the eigenvalue decomposition of ${\bm\Phi}_k$, where ${\mathbf{U}}_k \in \mathbb{C}^{N\times N}$ is a unitary matrix, and $\bm\Lambda_k$ is a diagonal matrix. Let ${\mathbf{u}}_{k,k'}$ denote the $k'$th column of ${\mathbf{U}}_k$, and it holds that ${\bm\phi}_{k,k}={\mathbf{U}}_k^{\mathsf{H}}{\bm\Lambda}_k{\mathbf{u}}_{k,k}$. Therefore, we have
\begin{align}
(\lambda{\mathbf{I}}+{\bm{\Phi}}_k)^{-1}{\bm\phi}_{k,k}&=(\lambda{\mathbf{U}}_k^{\mathsf{H}}{\mathbf{U}}_k+{\mathbf{U}}_k^{\mathsf{H}}{\bm\Lambda}_k{\mathbf{U}}_k)^{-1}{\mathbf{U}}_k^{\mathsf{H}}{\bm\Lambda}_k{\mathbf{u}}_{k,k}\nonumber\\
&={\mathbf{U}}_k^{\mathsf{H}}(\lambda{\mathbf{I}}+{\bm\Lambda}_k)^{-1}{\bm\Lambda}_k{\mathbf{u}}_{k,k},
\end{align}
and thus
\begin{align}
[{\mathbf{w}}_k]_n=\begin{cases}
{\mathbf{u}}_{k,k}^{\mathsf{H}}({\mathbf{I}}-(\lambda{\mathbf{I}}+{\bm\Lambda}_k)^{-1}{\bm\Lambda}_k){\mathbf{u}}_{k,k}& n=k,\\
-{\mathbf{u}}_{k,n}^{\mathsf{H}}(\lambda{\mathbf{I}}+{\bm\Lambda}_k)^{-1}{\bm\Lambda}_k{\mathbf{u}}_{k,k}& n\ne k.
\end{cases}
\end{align}
This means that
\begin{align}
{\mathbf{w}}_k&={\mathbf{U}}_k^{\mathsf{H}}{\mathbf{u}}_{k,k}-{\mathbf{U}}_k^{\mathsf{H}}(\lambda{\mathbf{I}}+{\bm\Lambda}_k)^{-1}{\bm\Lambda}_k{\mathbf{u}}_{k,k}\nonumber\\
&={\mathbf{U}}_k^{\mathsf{H}}({\mathbf{I}}-(\lambda{\mathbf{I}}+{\bm\Lambda}_k)^{-1}{\bm\Lambda}_k){\mathbf{u}}_{k,k}
\end{align}
Then, it has
\begin{subequations}
\begin{align}
&\int_{\mathcal{S}_{\mathrm{T}}}\left\lvert J_{k}(\mathbf{s})\right\rvert^2{\rm{d}}{\mathbf{s}}\nonumber\\
&\quad=\left\lvert\frac{a_{k,k}}{\lambda}\right\rvert^2{\mathbf{w}}_k^{\mathsf{H}}{\bm\Phi}_k{\mathbf{w}}_k=\left\lvert\frac{a_{k,k}}{\lambda}\right\rvert^2{\mathbf{w}}_k^{\mathsf{H}}{\mathbf{U}}_k^{\mathsf{H}}{\bm\Lambda}_k{\mathbf{U}}_k{\mathbf{w}}_k\\
&\quad=\left\lvert\frac{a_{k,k}}{\lambda}\right\rvert^2{\mathbf{u}}_{k,k}^{\mathsf{H}}({\mathbf{I}}-(\lambda{\mathbf{I}}+{\bm\Lambda}_k)^{-1}{\bm\Lambda}_k)\nonumber\\
&\qquad\qquad\qquad\qquad\cdot{\bm\Lambda}_k({\mathbf{I}}-(\lambda{\mathbf{I}}+{\bm\Lambda}_k)^{-1}{\bm\Lambda}_k){\mathbf{u}}_{k,k}.\label{monotone_function}
\end{align}
\end{subequations}

Then, we can prove that $\int_{\mathcal{S}_{\mathrm{T}}}\left\lvert J_{k}(\mathbf{s})\right\rvert^2{\rm{d}}{\mathbf{s}}$ is a monotone function with respect to $\lambda$, and so is $\sum_{k=1}^{K}\int_{\mathcal{S}_{\mathrm{T}}}\left\lvert J_{k}(\mathbf{s})\right\rvert^2{\rm{d}}{\mathbf{s}}$. 
\begin{IEEEproof}
Define the diagonal element of $\bm\Lambda_k$ as $\lambda_n$, and $\lambda_n\geq0$.
\eqref{monotone_function} can be denoted as 
\begin{subequations}
\begin{align}
\eqref{monotone_function}=f(\lambda)&=\left\lvert\frac{a_{k,k}}{\lambda}\right\rvert^2\sum_{n=1}^N \left\lvert[\mathbf{u}_{k,k}]_n\right\rvert^2 \left(1-\frac{\lambda_n}{\lambda+\lambda_n}\right)^2\lambda_n\\
&=\left\lvert a_{k,k}\right\rvert^2\sum_{n=1}^N \left\lvert[\mathbf{u}_{k,k}]_n\right\rvert^2 \frac{\lambda_n}{\left(\lambda+\lambda_n\right)^2}
\end{align}
\end{subequations}
\begin{align}
\frac{\partial f(\lambda)}{\partial \lambda}=-2\left\lvert a_{k,k}\right\rvert^2\sum_{n=1}^N \left\lvert[\mathbf{u}_{k,k}]_n\right\rvert^2 \frac{\lambda_n}{\left(\lambda+\lambda_n\right)^3}\leq 0
\end{align}
Therefore, $f(\lambda)$ is monotonically decreasing.
\end{IEEEproof}
Thus, we can use binary search to find $\lambda$, which is the solution to the following equation
\begin{align}
\sum_{k=1}^{K}\int_{\mathcal{S}_{\mathrm{T}}}\left\lvert J_{k}(\mathbf{s})\right\rvert^2{\rm{d}}{\mathbf{s}}=P.
\end{align}

Based on the solutions for the above variables $\{J_k({\mathbf{s}}),b_k,\epsilon_k,\beta_k,\eta_k\}_{k=1}^{K}$, problem \eqref{current_original} can be solved using a BCD procedure, which is summarized in \textbf{Algorithm~\ref{Algorithm1}}.

\begin{algorithm}[t!]
\caption{FP-based BCD Algorithm for WSSR Maximization Problem \eqref{current_original} in CAPA-based Communications}\label{Algorithm1}
\begin{algorithmic}[1]
    \State Initialize the iteration count $n=0$. 
    \State Initialize $J_k({\mathbf{s}})$ by using the MRT algorithm, i.e., $J_k^0({\mathbf{s}}) = \frac{H_k^*(\mathbf{s})\sqrt{P}}{\sqrt{\int_{\mathcal{S}_{\mathrm{T}}}\left\lvert H_{k}(\mathbf{s})\right\rvert^2{\rm{d}}{\mathbf{s}}}}$, and $b_k^0$ can be obtained.
    \Repeat
        \State Update $\epsilon_k^{n}$ by \eqref{epsilon_k};
        \State Update $\beta_k^{n}$ by \eqref{beta_k};
        \State Update $\eta_k^{n}$ by \eqref{eta_k};
        \State Update $J_k^{n}({\mathbf{s}})$ by \eqref{J_k_compact}, and $\lambda$ is obtained by binary search.
        \State Update $b_k^{n}$ by \eqref{b_k} and calculate the sum secure rate $(R^{\mathrm{sec}})^n$.
    \Until{the increment of $R^{\mathrm{sec}}$ becomes smaller than a predefined threshold.}
\end{algorithmic}
\end{algorithm}

\begin{figure}[t!]
    \centering
    \includegraphics[width=0.25\textwidth]{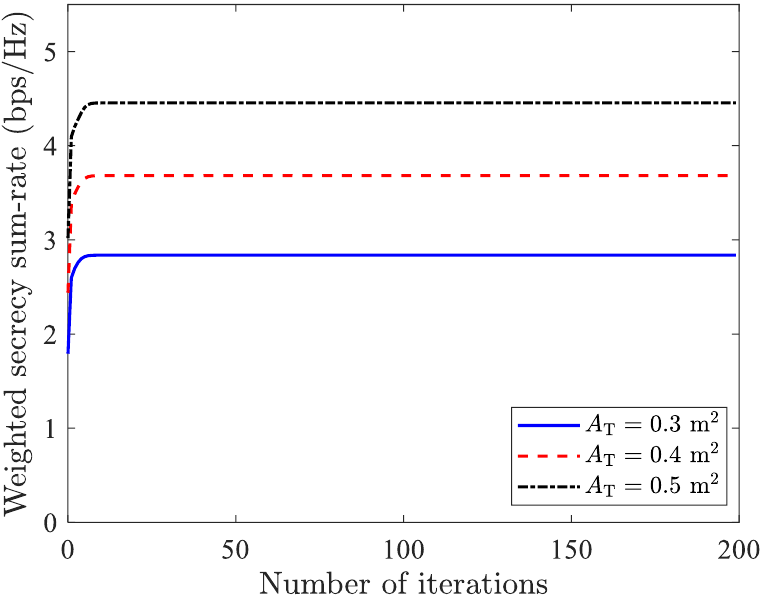} 
    \caption{Convergence of the proposed FP-based BCD algorithm.}
    \label{fig:convergence} 
\vspace{-15pt}
\end{figure}

\subsection{Convergence and Complexity}
We analyze the convergence of the proposed FP-based BCD algorithm as follows. The objective function in \eqref{current_transformation4_Objective} can be reformulated as a function of the variables $\{J_k({\mathbf{s}}),\epsilon_k,\beta_k,\eta_k\}_{k=1}^{K}$, denoted as $ f_{\text{obj}}\left(J_k({\mathbf{s}}),\epsilon_k,\beta_k,\eta_k\right)$. We omit $b_k$ since it is a non-negative constant that acts as an indicator function, ensuring the non-negativity of the objective function without affecting the convergence process of the algorithm.
The algorithm always satisfies the following relationship during iteration,
\begin{align}
f_{\text{obj}}&(J_k^{n+1}({\mathbf{s}}), \epsilon_k^{n+1}, \beta_k^{n+1},\eta_k^{n+1})\nonumber\\ 
&\overset{\text{(a)}}{\geq} f_{\text{obj}}(J_k^{n}({\mathbf{s}}), \epsilon_k^{n+1}, \beta_k^{n+1},\eta_k^{n+1})\nonumber\\
&\overset{\text{(b)}}{\geq}  f_{\text{obj}}(J_k^{n}({\mathbf{s}}), \epsilon_k^{n}, \beta_k^{n+1},\eta_k^{n+1})\nonumber\\
&\overset{\text{(c)}}{\geq}  f_{\text{obj}}(J_k^{n}({\mathbf{s}}), \epsilon_k^{n}, \beta_k^{n},\eta_k^{n+1})
\overset{\text{(d)}}{\geq}  f_{\text{obj}}(J_k^{n}({\mathbf{s}}), \epsilon_k^{n}, \beta_k^{n},\eta_k^{n})
\end{align}
where (a) holds because, when the variables other than $J_k({\mathbf{s}})$ are fixed, problem \eqref{only_current_optimize} becomes a convex optimization problem. The $J_k({\mathbf{s}})$ updated using the KKT condition satisfies the optimal solution.
While the (b), (c), and (d) hold because the updates for $\{\epsilon_k,\beta_k,\eta_k\}$ follow closed-form expressions, each being the optimal solution to a convex problem, as shown in \eqref{epsilon_k}, \eqref{beta_k} and \eqref{eta_k}. 
Fig. \ref{fig:convergence} further illustrates the average convergence performance under different transmitter aperture sizes through simulations.

We now present the complexity analysis of the FP-based BCD algorithm. According to \eqref{J_k_compact}, we observe that the main complexity arises from the integration operation involved in calculating the weighted correlation matrix $\{\bm{\Phi}_k\}_{k=1}^K$. To simplify the computation, let $c_{k,i}$ denote the $i$-th element of $\mathbf{c}_k \in \mathbb{R}^N$, and 
\begin{align}
{c}_{k,i}=\begin{cases}
\sqrt{b_{i}(1+\epsilon_{i})\left\lvert\eta_{i}\right\rvert^2}& ~i=1,\ldots,K,\\
\sqrt{\frac{1+\beta_k}{1+G_{\Gamma}}}& ~i=K+1,\ldots,N.
\end{cases}
\end{align}
Then, we have $\bm{\Phi}_k = \mathbf{c}_k\mathbf{c}_k^T \circ \mathbf{H}$, where $\mathbf{H}$ denotes the channel correlation matrix, and the $n$-th row and $k$-th column of $\mathbf{H}$ can be expressed as $[{\mathbf{H}}]_{n,k}=h_{n,k}=\int_{\mathcal{S}_{\mathrm{T}}}h_{n}({\mathbf{s}})h_{k}^*({\mathbf{s}}){\rm{d}}{\mathbf{s}}$, and 
\begin{align}
\begin{cases}
h_{k}({\mathbf{s}})=H_{k}({\mathbf{s}}), ~k=1,\ldots,K,\\
h_{K+q}({\mathbf{s}})=\bar{H}_{q}({\mathbf{s}}),~q=1,\ldots,Q.
\end{cases}
\end{align}
Therefore, once $\mathbf{H}$ is pre-computed, the calculation of $\{\bm{\Phi}_k\}_{k=1}^K$ during the algorithm's iteration no longer involves integration operations.
Additionally, by observing equations \eqref{phi} and \eqref{J_k_compact}, we find that the current pattern $J_{k}(\mathbf{s})$ is actually a linear combination of the spatial responses $\{h_n(\mathbf{s})\}_{n=1}^N$. Therefore, the integration operations in the update of optimization variables $\epsilon_k$, $\beta_k$, and $\eta_k$ are actually integrals between spatial responses, which can be obtained from $\mathbf{H}$. 
To summarize, the computational complexity of the proposed FP-based BCD algorithm primarily consists of two components. 
The first component involves computing the channel correlation matrix $\mathbf{H}$. By employing $M$-point Gauss-Legendre quadrature \cite{ref20}, this computation has a complexity of $\mathcal{O}(M^2N^2)$.
The second component corresponds to the matrix inversion required for computing $\{(\lambda{\mathbf{I}}+{\bm{\Phi}}_k)^{-1}\}_{k=1}^K$, with a computational complexity of $\mathcal{O}(LKN^3)$, where $L$ denotes the number of iterations. The Fourier-based discretization method in \cite{ref19} differs from the proposed approach and its computational complexity mainly arises from two parts: The first part involves the Fourier discretization of all channels, with a complexity of $\mathcal{O}(NN_FM^2)$. The second part comes from the matrix inversion during the update of the Fourier coefficients of the continuous current, with a complexity of $\mathcal{O}(LK N_F^3)$. Here, $N_F=\left(2\left\lceil \frac{L_x}{\lambda} \right\rceil +1\right)\left(2\left\lceil \frac{L_y}{\lambda} \right\rceil +1\right)$ represents the number of Fourier bases, where $L_x$ and $L_y$ denote the maximum projection lengths of aperture $\mathcal{S}_{\mathrm{T}}$. As the CAPA aperture increases and the frequency rises, $N_F$ will significantly increase. For instance, consider a CAPA with $L_x = L_y = 0.5 \, \text{m}$. When the frequency is 2.4 GHz, the number of Fourier bases $N_F$ is 81; at 7.8 GHz, $N_F$ is 729; and at 15 GHz, $N_F$ is 2601. Therefore, $N_F$ is typically $\gg N$, and the proposed method effectively reduces the computational complexity. Table \ref{TableI} presents the CPU runtime comparison for 10 iterations, conducted using MATLAB 2021a on an Intel i7-12700K processor.

\begin{table}[t!]
\vspace{-10pt}
\centering
\caption{Comparison of Average CPU Run Time.}\label{TableI}
\label{table:cpu_time}
\begin{tabular}{ccccccc}
\toprule
\text{Frequency} & \multicolumn{2}{c}{$A_T = 0.25 \, \mathrm{m}^2$} & \multicolumn{2}{c}{$A_T = 0.5 \, \mathrm{m}^2$} \\ 
\cmidrule(lr){2-3} \cmidrule(lr){4-5} 
 & Proposed & Fourier & Proposed & Fourier \\ 
\midrule
2.4 GHz & 1.49 s & 25.69 s & 1.32 s & 64.07 s \\ 
7.8 GHz & 1.35 s & 348.29 s & 1.25 s & 1031.67 s \\ 
\bottomrule
\end{tabular}
\vspace{-15pt}
\end{table}

\section{ZF Based Approach}
We also consider a zero-forcing (ZF) based scheme as a baseline. In this work, ZF beamforming is set to nullify inter-LUT interference. Under this circumstance, the beamforming vector for LUT $k$ must satisfy the following conditions:
\begin{subequations}\label{ZF_Condition}
\begin{align}
\int_{\mathcal{S}_{\mathrm{T}}}\bar{H}_{q}({\mathbf{s}}){J_k({\mathbf{s}})}{\rm{d}}{\mathbf{s}}&=0,~\forall q, \\
\int_{\mathcal{S}_{\mathrm{T}}}{{H}}_{i}({\mathbf{s}}){J_k({\mathbf{s}})}{\rm{d}}{\mathbf{s}}&=0,~\forall i\ne k,\\
\int_{\mathcal{S}_{\mathrm{T}}}{{H}}_{k}({\mathbf{s}}){J_k({\mathbf{s}})}{\rm{d}}{\mathbf{s}}&\ne0.
\end{align}
\end{subequations}
In this section, both $\{H_k({\mathbf{s}})\}_{k=1}^{K}$ and $\{\bar{H}_q({\mathbf{s}})\}_{q=1}^{Q}$ are in their non-normalized forms. An intuitive choice of $\{J_k({\mathbf{s}})\}_{k=1}^{K}$ that achieves zero-interference is given as follows.
\vspace{-5pt}
\begin{theorem}\label{theorem2}
Given $\{H_k({\mathbf{s}})\}_{k=1}^{K}$ and $\{\bar{H}_q({\mathbf{s}})\}_{q=1}^{Q}$, the following source current distribution satisfies the conditions in \eqref{ZF_Condition}:
\begin{align}
{{J}}_k^{\rm{ZF}}({\mathbf{s}})=\sum_{n=1}^{N}v_{k,n}h_{n}^*(\mathbf{s}),~N=K+Q,
\end{align}
where ${\mathbf{v}}_k=[v_{k,1},\ldots,v_{k,N}]^{T}$ is the $k$-th column of ${\mathbf{H}}^{-1}$. 
\end{theorem}
\vspace{-5pt}
\begin{IEEEproof}
We have
\begin{align}
\int_{\mathcal{S}_{\mathrm{T}}}{{h}}_{n}({\mathbf{s}}){J_k^{\rm{ZF}}({\mathbf{s}})}{\rm{d}}{\mathbf{s}}=\sum_{n'=1}^{N}v_{k,n'}h_{n,n'}=[h_{n,1},\ldots,h_{n,N}]{\mathbf{v}}_k.
\end{align}
By the definition of ${\mathbf{v}}_k$, we know that ${\mathbf{H}}^{-1}=[{\mathbf{v}}_1,\ldots,{\mathbf{v}}_N]$. Since ${\mathbf{H}}{\mathbf{H}}^{-1}={\mathbf{I}}$, we find 
\begin{align}
&\int_{\mathcal{S}_{\mathrm{T}}}{{h}}_{n}({\mathbf{s}}){J_k^{\rm{ZF}}({\mathbf{s}})}{\rm{d}}{\mathbf{s}}=[h_{n,1},\ldots,h_{n,N}]{\mathbf{v}}_k=0,~\forall n\ne k,\\
&\int_{\mathcal{S}_{\mathrm{T}}}{{h}}_{k}({\mathbf{s}}){J_k^{\rm{ZF}}({\mathbf{s}})}{\rm{d}}{\mathbf{s}}=[h_{k,1},\ldots,h_{k,N}]{\mathbf{v}}_k=1.
\end{align}
This completes the proof.
\end{IEEEproof}
By further considering the power allocation issue, we can design the current as follows:
\begin{align}\label{J_ZF}
{\hat{J}_k^{\rm{ZF}}({\mathbf{s}})}=\sqrt{P_k}\frac{{J}_k^{\rm{ZF}}({\mathbf{s}})}{\sqrt{\int_{\mathcal{S}_{\mathrm{T}}}\lvert {{J}}_k^{\rm{ZF}}({\mathbf{s}})\rvert^2{\rm{d}}{\mathbf{s}}}}.
\end{align}
Substituting \eqref{J_ZF} into problem \eqref{current_original}, the following power allocation problem is obtained:
\begin{subequations}\label{power_allocation}
\begin{align}
\max_{P_k}~&\sum_{k=1}^{K}\alpha_k\log_2\left(1+\frac{P_k}
{\sigma_k^2\int_{\mathcal{S}_{\mathrm{T}}}\lvert {{J}}_k^{\rm{ZF}}({\mathbf{s}})\rvert^2{\rm{d}}{\mathbf{s}}}\right)\\
{\rm{s.t.}}~&\sum_{k=1}^{K}P_k \leq P,
\end{align}
\end{subequations}
which can be solved via the water-filling method \cite{ref28}. The optimal solution to \eqref{power_allocation} is $P_k=\left(\mu \alpha_k-\sigma_k^2\int_{\mathcal{S}_{\mathrm{T}}}\lvert {{J}}_k^{\rm{ZF}}({\mathbf{s}})\rvert^2{\rm{d}}{\mathbf{s}}\right)^+$, where $\mu =\frac{P+\sum_{k=1}^C\sigma_k^2\int_{\mathcal{S}_{\mathrm{T}}}\lvert {{J}}_k^{\rm{ZF}}({\mathbf{s}})\rvert^2{\rm{d}}{\mathbf{s}}}{\sum_{k=1}^C\alpha_k}$ and $C$ is the number of non-zero $P_k$.

The water-filling method has a worst-case complexity of $\mathcal{O}(K)$. Additionally, the complexity of the proposed ZF approach arises from two components. The first is computing the channel correlation matrix $\mathbf{H}$, with a complexity of $\mathcal{O}(M^2N^2)$. The second is calculating the inverse of the matrix $\mathbf{H}$, as detailed in \textbf{Theorem~\ref{theorem2}}, with a worst-case complexity of $\mathcal{O}(N^3)$. For the Fourier-based ZF method, aside from the complexity introduced by the water-filling algorithm, its main computational costs are: the Fourier discretization of all channels, with complexity $\mathcal{O}(NN_FM^2)$, and the pseudo-inverse of the $N_F \times N$ discrete channel matrix, with complexity $\mathcal{O}(N^2N_F)$.

\section{Numerical Results}
In this section, we present numerical results obtained through Monte Carlo simulations to evaluate the performance of the proposed FP-based BCD algorithm and ZF-based method for CAPA secure beamforming. Unless stated otherwise, the following simulation parameters are used throughout the analysis. It is assumed that the CAPA transmitting surface, which is square-shaped, lies on the $xy$\text{-}plane and is centered at the origin of the coordinate system, defined as:  
\begin{equation}
\mathcal{S}_{\mathrm{T}} = \left\{ [s_x, s_y, 0]^T \, \middle| \, |s_x| \leq \frac{L_x}{2}, \, |s_y| \leq \frac{L_y}{2} \right\},
\end{equation}  
where $L_x = L_y = \sqrt{A_{\mathrm{T}}}$, and $A_{\mathrm{T}} = 0.25 ~\mathrm{m}^2$. The system contains $K = 8$ LUTs and $Q = 3$ Eves, which are randomly positioned within the following rectangular cuboid region:  
\begin{equation}
\mathcal{R} = \left\{ 
[r_x, r_y, r_z]^T
\middle|
\begin{array}{l}
|r_x| \leq U_x, \; |r_y| \leq U_y, \\
U_{z,\min} \leq r_z \leq U_{z,\max}
\end{array}
\right\},
\end{equation}  
where $U_x = U_y = 5~\mathrm{m}$, $U_{z,\mathrm{min}} = 15~\mathrm{m}$, and $U_{z,\mathrm{max}} = 30~\mathrm{m}$. Without loss of generality, assume that the polarization directions of all LUTs and Eves are aligned along the y-axis, i.e. $\hat{\mathbf{u}}_k = \hat{\mathbf{u}}_q =\hat{\mathbf{u}}_y = [0, 1, 0]^T$. The frequency of the source current is $f = 2.4 ~\mathrm{GHz}$, with the intrinsic impedance set to $\eta = 120\pi~ \Omega$. 
The maximum transmit power of BS is set to $P = 10 ~\mathrm{mA}^2$, while the noise power for LUTs and Eves is configured as $\sigma_k^2=\bar{\sigma}_q^2 = 5.6 \times 10^{-3} ~\mathrm{V}^2/\mathrm{m}^2$. Each LUT is treated equally, with the weight assigned as $\alpha_k = 1, \forall k$. The number of sampling points for the Gauss-Legendre quadrature method is set to $M = 10$. The proposed FP-based BCD algorithm initializes using MRT, i.e. $J_k(\mathbf{s})=\sqrt{P}\frac{{H}_k^*({\mathbf{s}})}{\sqrt{\int_{\mathcal{S}_{\mathrm{T}}}\lvert {{H}}_k({\mathbf{s}})\rvert^2{\rm{d}}{\mathbf{s}}}}$. All results are averaged over 200 independent random channel realizations unless specified otherwise.

We consider the following scheme as the benchmark:
\begin{itemize}
    \item \textbf{Fourier-based approach \cite{ref19}}: This method approximates the continuous current distribution as a finite sum of Fourier series, thereby discretizing the current distribution. The optimization problem of the function is transformed into an optimization problem of Fourier coefficients, which can then be solved using techniques commonly applied in conventional discretized MIMO systems. After obtaining the discrete Fourier coefficients of the source current, the continuous source current pattern can be reconstructed using the Fourier series.
    
    \item \textbf{Discrete MIMO}: The discrete MIMO array is arranged on the same area as the CAPA surface $\mathcal{S}_{\mathrm{T}}$, with antennas spaced according to $d = \frac{\lambda}{2}$. Therefore, the total number of discrete antennas is given by $N_{\rm MIMO} = \left\lceil \frac{L_x}{d} \right\rceil \times \left\lceil \frac{L_y}{d} \right\rceil$. The effective aperture area of each antenna is given by $A_d = \frac{\lambda^2}{4\pi}$. In the simulation, the coordinate position of the $(n_x, n_y)$-th discrete antenna is given  by
    \begin{equation}
        \bm{s}_{n_x, n_y} = \begin{bmatrix}
            (n_x - 1)d - \frac{L_x}{2}, \; (n_y - 1)d - \frac{L_y}{2}, \; 0
        \end{bmatrix}^T.
    \end{equation}
   Then, the channel between the $(n_x, n_y)$-th discrete antenna and LUT $k$, as well as Eve $q$ can be calculated as follows:
    \begin{equation}
        h_{(n_x, n_y),k} = \sqrt{A_d} \mathbf{\hat{u}}_k^T \mathbf{G}(\mathbf{r}_k, \mathbf{s}_{n_x, n_y}) \mathbf{\hat{u}}_y.
    \end{equation}
    \begin{equation}
        \bar{h}_{(n_x, n_y),q} = \sqrt{A_d} \mathbf{\hat{u}}_q^T \mathbf{G}(\mathbf{r}_q, \mathbf{s}_{n_x, n_y}) \mathbf{\hat{u}}_y.
    \end{equation}
By applying the same optimization reformulation process as in Section III, the WSSR maximization problem for the discrete precoding vectors is derived, which is referred to as MIMO-optimized. Additionally, the ZF and MRT methods can be applied to the discretized MIMO system, resulting in MIMO-ZF and MIMO-MRT, respectively, to reduce beamforming complexity~\cite{ref27}.
\end{itemize}

\begin{figure}[t!]
\vspace{-10pt}
    \centering
    \includegraphics[width=0.31\textwidth]{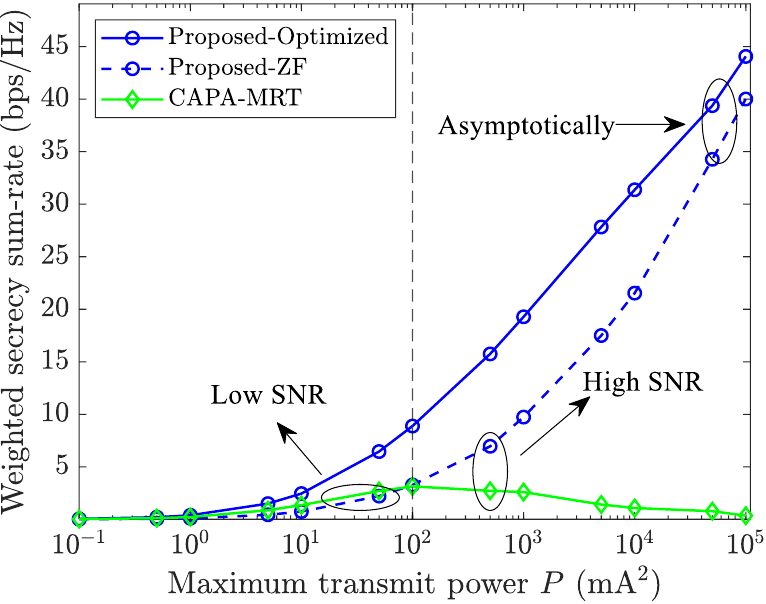} 
    \caption{WSSR for non-approximate beamforming schemes.}
    \label{fig:Proposedvs} 
\vspace{-10pt}
\end{figure}

\begin{figure}[t!]
\vspace{-1pt}
    \centering
    \includegraphics[width=0.31\textwidth]{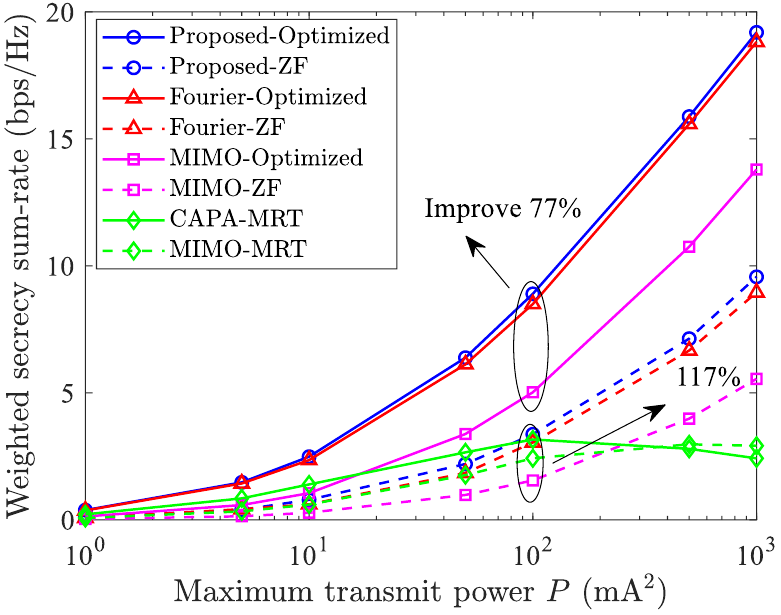} 
    \caption{WSSR versus transmit power.}
    \label{fig:P_T} 
\vspace{-15pt}
\end{figure}

\subsection{Comparison of WSSR for Non-Approximate Beamforming Schemes in CAPA Systems}
Fig. \ref{fig:Proposedvs} compares the performance of three non-approximate current distribution schemes in CAPA systems. As transmit power increases, the WSSR of the proposed ZF-based heuristic scheme approaches that of the FP-based BCD optimization method, as receiver noise becomes negligible at high power. By effectively suppressing interference between LUTs and eavesdropping by Eves, the proposed schemes achieve near-optimal secrecy rates.
Furthermore, the WSSR of the MRT algorithm exhibits an initial increase followed by a decrease. At low transmit power, MRT outperforms the ZF algorithm, but as power rises, ZF surpasses MRT. This occurs because higher power shifts the system from noise-dominated to interference-dominated, and MRT, which does not mitigate interference between LUTs and Eve, suffers from performance degradation.
Notably, when transmit power becomes excessively high (e.g., $P=10^5$ mA$^2$), transmission may need to be interrupted to prevent the leakage of confidential information.

\begin{figure}[t!]
\vspace{-15pt}
    \centering
    \includegraphics[width=0.31\textwidth]{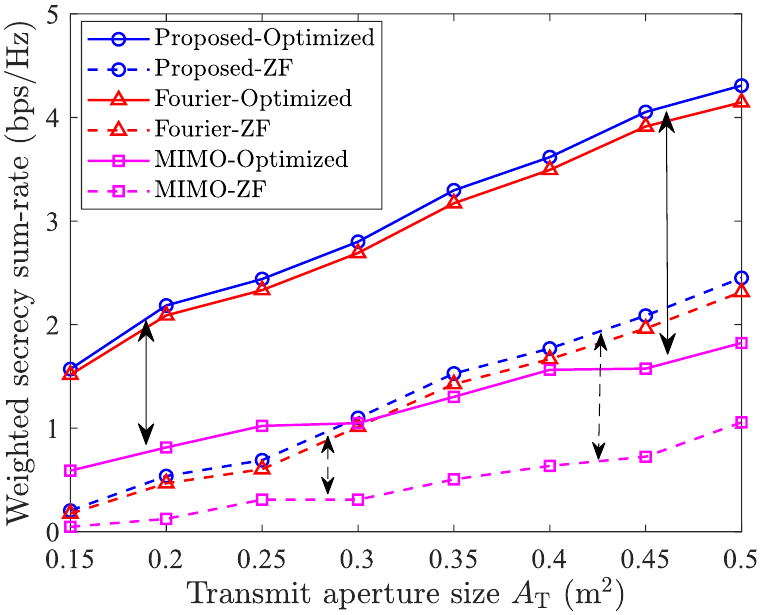} 
    \caption{WSSR versus transmit aperture size.}
    \label{fig:A_T} 
\vspace{-10pt}
\end{figure}
\subsection{WSSR Versus Transmit Power $P$}
Next, we present the performance gain of the CAPA system over the discrete MIMO system in secure communications. Fig. \ref{fig:P_T} illustrates the impact of transmit power on the WSSR. Firstly, it can be observed that, the WSSR of all methods, except for the MRT scheme, increases with the transmit power. However, CAPA-MRT gradually exhibits worse performance compared to MIMO-MRT. This is because MRT does not account for interference suppression at all, and CAPA, compared to discrete MIMO, not only increases the strength of the useful signal but also amplifies the interference signal, which ultimately results in poorer performance in interference-dominated systems.
Secondly, under all power conditions, the CAPA-based secure beamforming scheme demonstrates significant superiority over discrete MIMO methods, highlighting its advantage in enhancing WSSR. For instance, when $P = 10^2$ mA$^2$, the proposed optimization-based CAPA scheme achieves a 77\% improvement in WSSR compared to the optimization-based MIMO approach, while the proposed ZF-based CAPA scheme achieves a 117\% improvement over ZF-based MIMO. Specifically, at low transmit power, CAPA-MRT even outperforms the MIMO-optimized method. It proves that CAPA possesses better energy aggregation capability compared to discrete MIMO.
Furthermore, both the optimization-based and ZF-based proposed schemes outperform the Fourier-based approach, as they eliminate the approximation of continuous functions through Fourier series discretization, enabling direct acquisition of the optimal solution. As shown in Section III C, the computational complexity of the proposed scheme is also lower than that of the Fourier-based approach.

\subsection{WSSR Versus Aperture Size $A_{\rm T}$}
Fig. \ref{fig:A_T} investigates how the transmit aperture size $A_{\rm T}$ affects the WSSR performance. It can be seen that the WSSR of all methods improves as the aperture size increases. This is due to the increased spatial DoFs from a larger aperture, which enhances the beamforming gain. It effectively reduces inter-LUT interference and prevents information leakage. These benefits are comparable to the performance gains achieved by increasing the number of antennas in discrete MIMO systems. Furthermore, all methods under the CAPA system significantly outperform those under the discrete MIMO system for all aperture sizes. Additionally, as the aperture size increases, the WSSR gap between the two systems widens, demonstrating that the CAPA system can more fully exploit spatial DoFs.  Notably, when the aperture size $A_{\rm T}$ exceeds $0.3 \, \text{m}^2$, the WSSR achieved by the proposed ZF-based CAPA method surpasses that of the optimization-based MIMO method. Moreover, the proposed algorithm still shows a WSSR improvement compared to the Fourier-based method.

\begin{figure}[t!]
\vspace{-15pt}
    \centering
    \includegraphics[width=0.31\textwidth]{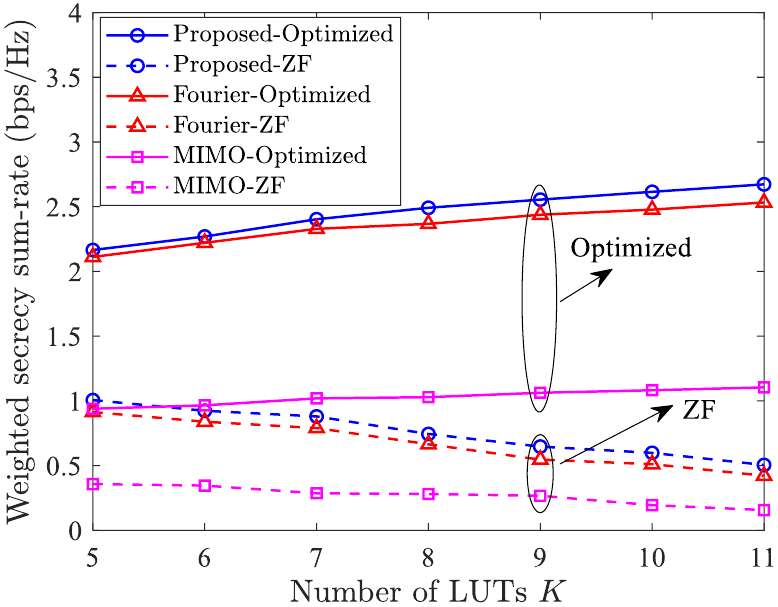} 
    \caption{WSSR versus number of LUTs.}
    \label{fig:LUTs} 
\vspace{-5pt}
\end{figure}

\begin{figure}[t!]
    \centering
    \includegraphics[width=0.31\textwidth]{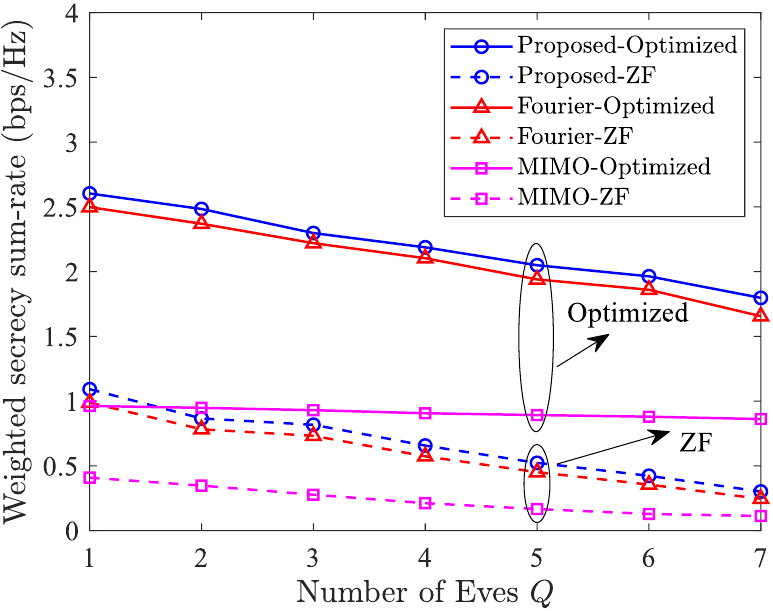} 
    \caption{WSSR versus number of Eves.}
    \label{fig:Eves} 
\vspace{-15pt}
\end{figure}

\subsection{WSSR Versus Number of LUTs $K$ and Eves $Q$}
Fig. \ref{fig:LUTs} illustrates the impact of the number of LUTs $K$ on WSSR performance. It can be observed that the proposed algorithm consistently outperforms the baseline schemes, with the CAPA system demonstrating significant advantages over the discrete MIMO system. As the number of LUTs increases, the optimization-based and ZF-based algorithms exhibit contrasting trends: the WSSR of the optimization-based algorithm improves, whereas that of the ZF-based algorithm declines. This decline is attributed to the increased number of LUTs, which weakens the orthogonality among the channels of all LUTs and Eves, making it more challenging for ZF to effectively suppress interference. In contrast, the optimization-based algorithm can more flexibly balance the impact of inter-LUT interference and information leakage on WSSR, enabling it to make more effective decisions.
Fig. \ref{fig:Eves} further investigates the trend of the WSSR as the number of Eves increases. It can be observed that the WSSR gradually decreases across all schemes as the number of Eves grows, which aligns with practical expectations, as more Eves inevitably lead to increased information leakage. Similarly, the performance of the CAPA system is significantly superior to that of the discrete MIMO system, and the proposed algorithms consistently outperform the benchmark schemes. 

\section{Conclusion}
This paper addressed the secure communication problem in CAPA systems with multiple LUTs and Eves. Simulation results highlighted the significant advantages of CAPA over discrete MIMO. We proposed an FP-based BCD optimization scheme to achieve optimal current distribution and a ZF-based heuristic scheme to reduce computational complexity while maintaining asymptotically optimal performance at high SNR. Compared to the Fourier-based discretization scheme, these methods enhanced WSSR performance and significantly reduced complexity. Additionally, the proposed continuous function inversion theory provided a robust tool for solving functional optimization problems, paving the way for future research on CAPA-based communication systems.

\appendices
\section{Proof of \textbf{Theorem~\ref{optimal_j_cov}}}\label{Appendix_F}
We utilize the CoV to establish \textbf{Theorem 1}. At any local maximum $\tilde{J}_k(\mathbf{s})$ of $g(J_{k}(\mathbf{s}))$, for any $\delta \to 0$, the following inequality holds:
\begin{equation}
    g(\tilde{J}_k(\mathbf{s})) \geq g(\tilde{J}_k(\mathbf{s}) + \delta U_k(\mathbf{s})),
\end{equation}
where $U_k(\mathbf{s})$ is any smooth function satisfying $U_k(\mathbf{s}) = 0,\forall \mathbf{s} \in \partial S_T $, and $\delta U_k(\mathbf{s})$ represents the variation of $J_k(\mathbf{s})$. The right-hand side of this inequality, defined as $\Psi_k(\delta) \triangleq g(\tilde{J}_k(\mathbf{s}) + \delta U_k(\mathbf{s}))$. To simplify the notation, let $A_i=b_i(1+\epsilon_i)|\eta_i|^2$, $B_k=b_k(1+\epsilon_k)\eta_k^*$, and $C_k=\frac{1+\beta_k}{1+G_{\Gamma}}$. Then, $\Psi_k(\delta)$ can be expressed based on \eqref{Lagrange_Function} as:
\begin{align}\label{Psi_delta}
\vspace{-15pt}
    \Psi_k&(\epsilon) = \delta^2\mathscr{A} + \delta \mathscr{B}+\mathscr{C}_k, 
\end{align}
where 
\begin{align}\label{A_Psi_delta}
\mathscr{A}&= \delta^2 \Bigg[ \sum_{i=1}^K A_i \Bigg| \int_{S_T} H_i(\mathbf{s}) U_k(\mathbf{s}) \, d\mathbf{s} \Bigg|^2 + \lambda \int_{S_T} |U_k(\mathbf{s})|^2 \, d\mathbf{s}\nonumber \\
    & +C_k\sum_{q=1}^Q \Bigg| \int_{S_T} \bar{H}_q^*(\mathbf{s}) U_k(\mathbf{s}) \, d\mathbf{s} \Bigg|^2  \Bigg], 
\end{align}
\begin{align}\label{B_Psi_delta}
\mathscr{B}&=2 \mathbb{R} \Bigg\{- B_k^* \int_{S_T} H_k^*(\mathbf{s}) U_k^*(\mathbf{s}) \, d\mathbf{s} \nonumber \\
    & +\sum_{i=1}^K A_i \int_{S_T} \int_{S_T} H_i(\mathbf{s}') \tilde{J}_k(\mathbf{s}') H_i^*(\mathbf{s}) U_k^*(\mathbf{s}) \, d\mathbf{s}'d\mathbf{s} \nonumber \\
    & +C_k\sum_{q=1}^Q \int_{S_T} \int_{S_T} \bar{H}_q(\mathbf{s}') \tilde{J}_k(\mathbf{s}') \bar{H}_q^*(\mathbf{s}) U_k^*(\mathbf{s}) \, d\mathbf{s}'d\mathbf{s} \nonumber \\
    & + \lambda \int_{S_T} \tilde{J}_k(\mathbf{s}) U_k^*(\mathbf{s}) \, d\mathbf{s} \Bigg\},
\end{align}
and $\mathscr{C}_k$ is a constant independent of $\delta$.
Since $g(J_k(\mathbf{s}))$ achieves its local maximum at $\tilde{J}_k(\mathbf{s})$, it follows that $\Psi_k(\delta)$) attains its maximum at $\delta = 0$, leading to the condition:
\begin{equation}\label{d_delta}
    \frac{d \Psi_k(\delta)}{d\delta} \bigg|_{\delta=0} = 0.
\end{equation}

By substituting \eqref{Psi_delta} into \eqref{d_delta}, the condition becomes:
\begin{equation}\label{COV_lemma}
    \mathbb{R} \left\{ \int_{S_T} U_k^*(\mathbf{s}) V_k(\mathbf{s}) \, d\mathbf{s} \right\} = 0, 
\end{equation}
where
\begin{align}
&V_k(\mathbf{s})= \sum_{i=1}^K A_i H_i^*(\mathbf{s}) \int_{S_T} H_i(\mathbf{s}') \tilde{J}_k(\mathbf{s}') \, d\mathbf{s}' \nonumber \\
&+C_k\sum_{q=1}^Q  \bar{H}_q^*(\mathbf{s}) \int_{S_T} \bar{H}_q(\mathbf{s}') \tilde{J}_k(\mathbf{s}') \, d\mathbf{s}'-B_k^* H_k^*(\mathbf{s})+\lambda \tilde{J}_k(s).
\end{align}

The equation \eqref{COV_lemma} must hold for any arbitrary $U_k(\mathbf{s})$. According to \textbf{Lemma~3} in \cite{ref20}, we have:
\begin{equation}
    V_k(\mathbf{s}) = 0, \quad \forall \mathbf{s} \in S_T,
\end{equation}
then, we have got the optimal current structure in \eqref{optimal_J_structure}.

\section{Proof of \textbf{Theorem~\ref{theorem1}}}\label{Appendix_E}
According to \eqref{Inversion_Operator}, we obtain
\begin{equation}\label{Theorem11}
\begin{split}
\int_{\mathcal{S}_{\mathrm{T}}}&\hat{\Psi}({\mathbf{s}}'',{\mathbf{s}}){\Psi}({\mathbf{s}},{\mathbf{s}}'){\rm{d}}{\mathbf{s}}=
\delta({\mathbf{s}}''-{\mathbf{s}}')+\sum_{i'=1}^{I}\psi_{i'}^*({\mathbf{s}}'')\psi_{i'}({\mathbf{s}}')\\
&-\sum_{i=1}^{I}\sum_{i'=1}^{I}\psi_{i}^*({\mathbf{s}}'')\psi_{i'}({\mathbf{s}}')[({\mathbf{I}}+{\bm{\Psi}})^{-1}]_{i,i'}\\
&-\sum_{i=1}^{I}\sum_{i'=1}^{I}\sum_{i''=1}^{I}\Bigg(\psi_{i}^*({\mathbf{s}}'')[({\mathbf{I}}+{\bm{\Psi}})^{-1}]_{i,i'}\psi_{i''}({\mathbf{s}}')\\
&\quad\quad \quad \quad \quad\quad \quad \quad\cdot\int_{\mathcal{S}_{\mathrm{T}}}\psi_{i'}({\mathbf{s}})\psi_{i''}^*({\mathbf{s}}){\rm{d}}{\mathbf{s}}\Bigg),
\end{split}
\end{equation}
which yields
\begin{equation}\label{Theorem12}
\vspace{-15pt}
\begin{split}
&\int_{\mathcal{S}_{\mathrm{T}}}\hat{\Psi}({\mathbf{s}}'',{\mathbf{s}}){\Psi}({\mathbf{s}},{\mathbf{s}}'){\rm{d}}{\mathbf{s}}=
\delta({\mathbf{s}}''-{\mathbf{s}}')+\sum_{i'=1}^{I}\psi_{i'}^*({\mathbf{s}}'')\psi_{i'}({\mathbf{s}}')\\
&-\sum_{i=1}^{I}\sum_{i'=1}^{I}\psi_{i}^*({\mathbf{s}}'')\psi_{i'}({\mathbf{s}}')[({\mathbf{I}}+{\bm{\Psi}})^{-1}]_{i,i'}\\
&-\sum_{i=1}^{I}\sum_{i''=1}^{I}\Bigg(\psi_{i}^*({\mathbf{s}}'')
\psi_{i''}({\mathbf{s}}')\\
&\quad\quad \quad \quad \quad\quad\cdot\sum_{i'=1}^{I}\int_{\mathcal{S}_{\mathrm{T}}}\psi_{i'}({\mathbf{s}})\psi_{i''}^*({\mathbf{s}}){\rm{d}}{\mathbf{s}}[({\mathbf{I}}+{\bm{\Psi}})^{-1}]_{i,i'}\Bigg).
\end{split}
\end{equation}
It follows
\begin{equation}\label{First_Equality_Transformation}
\begin{split}
&\int_{\mathcal{S}_{\mathrm{T}}}\hat{\Psi}({\mathbf{s}}'',{\mathbf{s}}){\Psi}({\mathbf{s}},{\mathbf{s}}'){\rm{d}}{\mathbf{s}}=
\delta({\mathbf{s}}''-{\mathbf{s}}')\\
&\quad+\sum_{i=1}^{I}\sum_{i'=1}^{I}\psi_{i}^*({\mathbf{s}}'')\psi_{i'}({\mathbf{s}}')(\delta_{i,i'}\\
&\quad-[({\mathbf{I}}+{\bm{\Psi}})^{-1}]_{i,i'}-\sum_{i''=1}^{I}[({\mathbf{I}}+{\bm{\Psi}})^{-1}]_{i,i''}[{\bm{\Psi}}]_{i'',i'}),
\end{split}
\end{equation}

where $\delta_{i,i'}$ denotes the Kronecker delta. It is worth noting that $\delta_{i,i'}
-[({\mathbf{I}}+{\bm{\Psi}})^{-1}]_{i,i'}-\sum_{i''=1}^{I}[({\mathbf{I}}+{\bm{\Psi}})^{-1}]_{i,i''}
[{\bm{\Psi}}]_{i'',i'}$ is the $(i,i')$-th element of the following matrix:
\begin{align}
&{\mathbf{I}}-({\mathbf{I}}+{\bm{\Psi}})^{-1}-({\mathbf{I}}+{\bm{\Psi}})^{-1}{\bm{\Psi}}\nonumber\\
&\quad={\mathbf{I}}-({\mathbf{I}}+{\bm{\Psi}})^{-1}({\mathbf{I}}+{\bm{\Psi}})={\mathbf{I}}-{\mathbf{I}}={\mathbf{0}}.
\end{align}

This, together with \eqref{First_Equality_Transformation}, yields
\begin{equation}\label{First_Equality_Proof}
\int_{\mathcal{S}_{\mathrm{T}}}\hat{\Psi}({\mathbf{s}}'',{\mathbf{s}}){\Psi}({\mathbf{s}},{\mathbf{s}}'){\rm{d}}{\mathbf{s}}=
\delta({\mathbf{s}}''-{\mathbf{s}}')
\end{equation}
Following similar derivation steps, we can show
\begin{align}
\int_{\mathcal{S}_{\mathrm{T}}}{\Psi}({\mathbf{s}}'',{\mathbf{s}})\hat{\Psi}({\mathbf{s}},{\mathbf{s}}'){\rm{d}}{\mathbf{s}}=\delta({\mathbf{s}}''-{\mathbf{s}}'),
\end{align}
which completes the proof.

\newpage

%
%
%
%

\vfill

\end{document}